\documentclass[prd,showkeys,showpacs,epsfig,nofootinbib,floatfix]{revtex4} 

\usepackage[dvips]{graphicx} 
\usepackage{amssymb}

\newcommand{\be}{\begin{equation}}
\newcommand{\ee}{\end{equation}}
\newcommand{\bea}{\begin{eqnarray}}
\newcommand{\eea}{\end{eqnarray}}
\newcommand{\refeq}[1]{equation~(\ref{eq:#1})}          
\newcommand{\refeqs}[2]{equations~(\ref{eq:#1})--(\ref{eq:#2})}          

\newcommand{\reffig}[1]{Fig.~\ref{fig:#1}}
\newcommand{\refsec}[1]{section~\ref{sec:#1}}          
\newcommand{\refSec}[1]{Section~\ref{sec:#1}}          

\renewcommand{\v}[1]{\mathbf{#1}}

\newcommand{\vx}{\v{x}}

\newcommand{\vnhat}{\v{\hat{n}}}
\renewcommand{\d}{\delta}
\renewcommand{\k}{\kappa}

\renewcommand{\l}{\ell}
\newcommand{\fSky}{f_{{\rm Sky}}}

\newcommand{\iMpch}{\:h/{\rm Mpc}}
\newcommand{\Phim}{\Phi_-}
\newcommand{\Om}{\Omega_m}
\newcommand{\OL}{\Omega_\Lambda}
\newcommand{\DPhim}{D_{\Phim}}
\newcommand{\zbar}{\bar{z}}
\newcommand{\erfc}{{\rm erfc}}

\newcommand{\fR}{f(R)}
\renewcommand{\P}{\mathcal{P}}

\begin{document}

\title{Weak Lensing Probes of Modified Gravity}

\author{Fabian Schmidt}
\affiliation{Department of Astronomy \& Astrophysics, The University of
Chicago, Chicago, IL 60637-1433}
\affiliation{Kavli Institute for Cosmological Physics, Chicago, IL 
60637-1433}
\email{fabians@oddjob.uchicago.edu}

\begin{abstract}
We study the effect of modifications to General Relativity on large scale weak lensing 
observables. In particular, we consider three modified gravity scenarios: 
$f(R)$ gravity, the DGP model, and TeVeS theory. 
Weak lensing is sensitive to the growth of structure and the relation
between matter and gravitational potentials, both of which will in general be 
affected by modified gravity. 
Restricting ourselves to linear scales, we compare the predictions for 
galaxy-shear and shear-shear correlations of each modified gravity cosmology 
to those of an effective Dark Energy cosmology with the same expansion history.
In this way, the effects of modified gravity on the growth of perturbations are
separated from the expansion history. We also propose a test which isolates the 
matter-potential relation from the growth factor and matter power spectrum.
For all three modified gravity models, the predictions for galaxy and 
shear correlations will be discernible from those of Dark Energy with very 
high significance in future weak lensing surveys. Furthermore, each model
predicts a measurably distinct scale dependence and redshift evolution
of galaxy and shear correlations, which can be traced back to the physical
foundations of each model.
We show that the signal-to-noise for 
detecting signatures of modified gravity is much higher for weak lensing 
observables as compared to the ISW effect, measured via the galaxy-CMB 
cross-correlation.
\end{abstract}


\date{\today}

\maketitle

\section{Introduction}
\label{sec:intro}

A number of independent observations, ranging from Supernovae to the 
cosmic microwave background (CMB) and its cross-correlation with foreground 
galaxies \cite{RiessEtal,KowalskiEtal,WMAP5,Giannantonio,Pietrobon}, 
have now firmly
established that the expansion of the Universe is accelerating, and that a purely 
matter-dominated universe is not consistent with measurements. Commonly,
these observations are ascribed to an additional smooth stress-energy
component, ``Dark Energy'' (DE), pervading the Universe \cite{FriemanEtal08}.
However, instead of attributing the accelerated expansion to our lack
of understanding of the constituents of the Universe, one can take the
alternative approach and attribute it to our lack of understanding of
gravity on cosmological scales. For example, an effective weakening
of gravity on the largest scales could explain the accelerated expansion
and other cosmological observations without invoking an additional smooth
``dark'' component. Several such modified gravity scenarios have
been proposed. The difficulty in constructing a consistent
theory of gravity which preserves the success of General Relativity (GR)
in the Solar System and the early Universe has however limited the number of
proposed theories to only a handful.

Seen from a fundamental physics point of view, the two approaches of 
General Relativity coupled with a smooth Dark Energy
(GR+DE) and modified gravity are apparently quite distinct. However,
there is enough freedom in both scenarios to match any given expansion 
history of the Universe. Thus, the expansion
history, measured primarily with Supernovae and the CMB, is not sufficient
to distinguish between these two ``fundamentally different'' scenarios.

Going beyond the smooth background Universe offers a multitude of additional
observables probing the evolution of structure formation in the Universe.
For the purpose of constraining modified gravity, the evolution
of the cosmological gravitational potentials, and their relation to the matter
overdensities and velocities are the most sensitive probes 
\cite{JainZhang,Caldwell,Amendola,Bertschinger}. Several such 
observational tests of gravity have
been proposed: the \textit{Integrated Sachs-Wolfe (ISW)} effect 
\cite{SachsWolfe,Afshordi,ZhangfR,SongEtalDGP,Schmidt07} directly probes the evolution 
of the potentials on horizon-size scales. While quite sensitive as a probe
of gravity, it is restricted to
large scales, and offers only a limited amount of signal-to-noise. 
The \textit{matter power spectrum} probes the growth of matter
perturbations on a range of scales. While it can be measured to high 
precision and to very small scales, and hence 
offers a large amount of information, it is affected by the non-linearities
of structure formation, and galaxy bias, which are not well understood
for modified gravity models. A cleaner method of probing gravity with
galaxies is using \textit{velocity correlations} \cite{ZhangEtal,JainZhang,SongKoyama}. 
In principle, these tests can isolate signatures of modified gravity.
However, they will be challenging to perform observationally.

Several deep, wide-field galaxy surveys are in the planning stage which are designed
to measure galaxy counts and weak lensing shear with unprecedented accuracy.
Galaxy and shear correlations can be used to measure the expansion history
and growth of structure in the Universe to high precision. Additionally,
the potential of weak lensing observables to probe gravity has been
shown in \cite{KnoxSongTyson,Song2005,Song2006,JainZhang,Tsujikawa2008}.
In this paper, we investigate the constraints on modified gravity theories to
be expected from galaxy and shear correlations in future surveys. 
Since weak lensing is caused by the potential wells along the line of sight, 
it can be used to measure the scale dependence and cosmological evolution of
these potentials. In addition, correlating foreground galaxies with the
shear of background galaxies offers a test of the matter-potential relation.
We will show that future weak lensing surveys will place very tight 
constraints on gravity. In other words, 
they will perform stringent tests of the smooth Dark Energy scenario, which 
will have to match all of the weak lensing observables in order to 
remain a viable model.

Regarding the potential of these precision tests,
it is important to keep several caveats in mind: first, the non-linear
evolution in modified gravity models is not yet understood in a realistic
cosmological context. Hence, in this paper we will limit ourselves to scales 
where linear theory is appropriate. Second, correlations involving galaxy
number counts are affected by the galaxy bias, which is a priori unknown.
We will point out how the effects of bias can hopefully be disentangled
from modified gravity effects. Third, it is always 
possible to find a general, non-smooth Dark Energy model which mimics the 
predictions of any given
modified gravity model \cite{HuSawicki07,Bertschinger2}; note that such a 
DE model will have significant density and anisotropic stress perturbations on
sub-horizon scales. 

In this paper, we will study three different modified gravity scenarios, 
focusing on their predictions for lensing 
observables: (i) $\fR$ gravity \cite{CarrollEtal,Carroll2,Mota2006,NoOd2006,SongEtal07}; 
(ii) the Dvali-Gabadadze-Porrati (DGP)
model \cite{DGP,Deffayet01,DeffayetEtal02}; and 
(iii) Tensor-Vector-Scalar theory (TeVeS) \cite{Bekenstein}. While the first two
models are able to achieve accelerated expansion without dark energy,
TeVeS is designed to explain observations without dark matter by showing
MOND-behavior in certain regimes. These three theories encompass very
different approaches to the problem of constructing a consistent theory
of gravity.
Hence, it is interesting to determine not only whether observations can detect
departures from GR, but also whether they would be able to discriminate 
among different modified gravity scenarios.

In addition to predicting a late-time accelerated expansion of the Universe,
all these models have been designed to approach GR in the high 
curvature regime, which applies to the Solar System as well as the early 
Universe. In this way, the models pass Solar System tests while making
predictions close to the standard cosmological model for the CMB and Big Bang 
nucleosynthesis. 

Weak lensing probes the growth of large-scale structure which is sensitive
to the background expansion history, as the expansion rate enters the 
evolution equations of perturbations. Since two of the models discussed
here (DGP and TeVeS) predict expansion histories somewhat different from
$\Lambda$CDM, we compare them with smooth Dark Energy models designed to 
mimic the expansion history of these models by employing a suitable equation
of state $w(a)$. Deviations in the weak lensing 
predictions of DGP and TeVeS from the corresponding dark energy model are 
then solely due to modified gravity effects on the growth of structure
and the matter-potential relation, and independent of the 
exact expansion history.

The structure of the paper is as follows. In \refSec{WL}, general expressions
for galaxy and shear correlations applicable to modified gravity are given, 
and we point out out the different channels through which modified gravity 
can affect the observables. We also present a cosmological test of the Poisson
equation using weak lensing and galaxy correlations.
In \refsec{MG}, we briefly describe the modified gravity
models studied here, and outline the characteristics which determine their
lensing predictions. We then present forecasts for galaxy-shear and
shear-shear correlations in future surveys in \refsec{res}, showing the 
deviations of the modified gravity predictions from those of GR+DE, as well
as the expected signal-to-noise in actual surveys.
We conclude in \refsec{concl}.

\section{Weak Lensing Correlations}
\label{sec:WL}

\subsection{Weak lensing in modified gravity}

In this section, we summarize the lensing observables considered in this
paper, generalizing the standard expressions to the case of modified
gravity. We assume a flat Universe throughout. Our metric convention is
\cite{ModCosm}:
\be
ds^2 = a^2(\eta) \left [ -(1+2\Psi)d\eta^2 + (1+2\Phi) d\v{x}^2 \right ],
\label{eq:metric}
\ee
where $a(\eta)$ is the scale factor, $\eta$ denotes conformal time,
and $\Phi$, $\Psi$ are the cosmological potentials. In General Relativity
(GR), $\Phi=-\Psi$ in the absence of anisotropic stress, which is the case 
at late times in $\Lambda$CDM and most dark energy models. 
In addition, the potentials
are conventionally re-expressed in terms of the matter overdensity
using the Poisson equation. 
In case of modified gravity, both of these assumptions do not necessarily hold.
In the following we give generalized expressions for the different
observables which apply to the case of modified gravity 
(see also \cite{JainZhang,KnoxSongTyson}).
We restrict ourselves to linear perturbations valid on large scales
($k \lesssim 0.1\iMpch$). Since the $\fR$ and DGP models recover the GR+DE
limit at early times in the matter-dominated epochs, it is convenient to
refer observables to the matter overdensity at high redshift $z_m$. For
definiteness, we take $z_m=50$. Then, for any viable $f(R)$ and DGP model,
$P(k,z_m)$ is identical to that expected in GR.

Weak lensing surveys use the observed ellipticities of galaxies
to reconstruct a map of the cosmic shear, which can then be used to
infer the convergence $\kappa$ \cite{BartSchneider,HuJain}. 
In the following, we will use the terms convergence and shear interchangeably,
understanding that they are reconstructed from observed galaxy ellipticities.
Using standard approximations, the convergence in the direction $\vnhat$,
given by the Laplacian of the projected lensing potential, is expressed as
a line of sight integral to the source at redshift $z_s$:

\be
\kappa(\vnhat) = - \int_0^{z_s} \frac{dz}{H(z)} \nabla^2
\Phim(\vx(z),z) W_L(\chi_s,\chi(z)),
\label{eq:kappa}
\ee
where $\chi(z)$ is the comoving distance out to redshift $z$,
$H(z) = \frac{da}{dt}/a$ is the expansion rate,
$\Phim \equiv 1/2(\Phi-\Psi)$, $\nabla^2$ denotes the Laplacian in terms 
of comoving coordinates, and the lensing weight function is given by:
\be
W_L(\chi_s,\chi) = \frac{\chi}{\chi_s}(\chi_s-\chi)
\label{eq:W_L}
\ee
Hence, the convergence is sensitive to the potential $\Phim$ in a range of distances
centered around $\chi \sim\chi_s/2$. Note that in the case of GR and smooth
Dark Energy (referred to as ``GR+DE'' in the following), 
$\Phim = \Phi=-\Psi$, and we can apply the Poisson equation:
\be
\nabla^2 \Phim \stackrel{{\rm GR+DE}}{=} 
4\pi G a^2 \rho_m\:\d = \frac{3}{2}\Om H_0^2 a^{-1}\d
\label{eq:PoissonLCDM}
\ee
In modified gravity, this equation does not hold anymore, and we define
a generalized ``Poisson factor'' $\DPhim$ relating the potential $\Phim$
to the matter overdensity $\d$ in Fourier space:
\be
\DPhim \equiv \frac{\Phim(k,z)}{\d(k,z)} \stackrel{{\rm GR+DE}}{=} 
\frac{3}{2}\Om \frac{H_0^2}{k^2} a^{-1}
\label{eq:Pratio}
\ee
Apparently, a departure of the Poisson factor $\DPhim$ from its expected value
in General Relativity
is a signature of the modification of the Poisson equation in alternative 
gravity theories (for a discussion of general, non-smooth Dark Energy models,
see \refsec{DE}). In addition, the growth of potential and matter
perturbations is affected by modified gravity. The growth factor of
matter, which we normalize at $z=z_m=50$ so that 
$D_m(k,z) \equiv \d(k,z)/\d(k,z=z_m)$, becomes scale-dependent in many
modified gravity models. We can combine the growth of structure
and the Poisson equation to obtain:
\be
\frac{\Phim(k,z)}{\d(k,z=z_m)} = \DPhim(k,z)\:D_m(k,z).
\ee
This allows us to express the galaxy-galaxy, galaxy-shear and shear-shear 
correlation coefficients, $C^{gg}(\l)$, $C^{g\k}(\l)$, $C^{\k\k}(\l)$ 
as follows:
\bea
C^{g_ig_j}(\l) &=& \frac{2}{\pi}\int dk\:k^2 P(k,z_m) 
\int dz\: b_iW_{g_i}(z) D_m(k,z) j_\l(k\chi(z))
\int dz'\: b_jW_{g_j}(z') D_m(k,z') j_\l(k\chi(z')) \label{eq:CggEx}\\
C^{g_i\k_j}(\l) &=& \frac{2}{\pi}\int dk\:k^2 P(k,z_m) 
\int dz\: b_iW_{g_i}(z) D_m(k,z) j_\l(k\chi(z))
\int dz'\: W_{\k_j}(z') k^2 \DPhim(k,z') D_m(k,z') j_\l(k\chi(z'))\quad \label{eq:CgkEx}\\
C^{\k_i\k_j}(\l) &=& \frac{2}{\pi}\int dk\:k^2 P(k,z_m) 
\int dz\: W_{\k_i}(z) k^2 \DPhim(k,z) D_m(k,z) j_\l(k\chi(z)) \nonumber\\
& & \times
\int dz'\: W_{\k_j}(z') k^2 \DPhim(k,z') D_m(k,z') j_\l(k\chi(z')) \label{eq:CkkEx}
\eea
Here, $P(k,z_m)$ is the matter power spectrum
at early times $z=z_m$, which in the case of the $\fR$ and DGP models is the same
as for a GR + smooth DE model with the same expansion history. The indices
$i$, $j$ denote different redshift bins, i.e.,
$W_{g_i}(z)$  is the galaxy redshift distribution for bin $i$, normalized to 1,
$b_i$ is the galaxy bias for the same bin,
and the shear weighting function of redshift bin $j$, $W_{\k_j}$, is given by:
\be
W_{\k_j}(z) = \frac{1}{H(z)}\int_z^{\infty} dz_s W_L(\chi(z_s),\chi(z)) W_{g_j}(z_s)
\ee
For $\l\gtrsim 10$, we can apply the Limber approximation \cite{Limber},
which simplifies the expressions considerably:
\bea
C^{g_ig_j}(\l) &=& \int dz \frac{H(z)}{\chi^2(z)} b_i W_{g_i}(z) b_j W_{g_j}(z)
\left [ D_m^2(k,z)\:P(k,z_m) \right ]_{k=\frac{l+1/2}{\chi(z)}}
\label{eq:Cgg} \\
C^{g_i\k_j}(\l) &=& \int dz \frac{H(z)}{\chi^2(z)} b W_{g_i}(z) W_{\k_j}(z)
\left [ \DPhim(k,z)\:D^2_m(k,z)\:k^2 P(k,z_m) \right ]_{k=\frac{l+1/2}{\chi(z)}}
\label{eq:Cgk} \\
C^{\k_i\k_j}(\l) &=& \int dz \frac{H(z)}{\chi^2(z)} W_{\k_i}(z) W_{\k_j}(z)
\left [ \DPhim^2(k,z)\:D^2_m(k,z)\:k^4 P(k,z_m) \right ]_{k=\frac{l+1/2}{\chi(z)}}
\label{eq:Ckk}
\eea
Weak lensing correlations thus depend on cosmology through three distinct
channels: (i) the background expansion history, via $H(z)$ and 
$W_L(\chi_s,\chi(z))$; (ii) the growth of perturbations and the Poisson 
equation, through $D_m(k,z)$ and $\DPhim(k,z)$; and (iii) the matter power 
spectrum at early times, $P(k,z_m)$. In case General Relativity holds
and the accelerated expansion is due a smooth Dark Energy, only the expansion 
history (i) is affected by the Dark Energy, i.e. modified from $\Lambda$CDM.
In case of the late-time acceleration modified gravity models $\fR$ and DGP,
both channels (i) and (ii) are affected. And the TeVeS model without dark 
matter modifies (i), (ii), as well as (iii) from the $\Lambda$CDM case. 

In this paper, we want to
illuminate how weak lensing correlations are sensitive to these three channels,
both in redshift evolution and in scale dependence. Clearly, it is
crucial to separate these effects in order to distinguish between GR+DE
and modified gravity; in addition, testing $D_m(k,z)$, $\DPhim(k,z)$, and
$P(k,z_m)$ separately will allow to distinguish between different modified
gravity scenarios.

The galaxy-shear correlation is proportional to the a priori
unknown bias of the foreground galaxies. Assuming that the bias is 
scale-independent on linear scales, one can marginalize over this
parameter for a given galaxy sample. An alternative is to consider the reduced 
galaxy-shear correlation $R^{g\k}(\l)$, defined by:
\be
R^{g_i\k_j}(\l) \equiv \frac{C^{g_i\k_j}(\l)}{\sqrt{C^{g_ig_i}(\l)}}
\label{eq:Rgk}
\ee
For redshift bins that are not too wide, the linear galaxy bias drops out
of this expression, so that $R^{g\k}$ is independent of $b_i$. Since it
is not straightforward to compute the errors on this quantity,
we will for simplicity only consider the galaxy-shear correlation divided by the bias, 
$C^{g\k}(\l)/b$, in this paper. In any case, we do not expect the constraints to 
be degraded very significantly when considering $R^{g\k}$ instead of $C^{g\k}$.

On small scales $k \sim 0.1\iMpch$ and above, the density contrast grows
to order unity, and non-linearities in the formation of structure become
important. Furthermore, all of the modified gravity theories studied here 
employ non-linear equations of motion which are expected to restore gravity
to General Relativity in high-density environments 
\cite{NavarroAcoleyen,HuSawickifR,DeffayetEtalDGPNL,Bekenstein}.
These mechanisms have been studied in special cases, e.g. spherically symmetric
and/or static cases \cite{NavarroAcoleyen,Bekenstein}; however, the details of 
cosmological structure 
formation in the non-linear regime have not been studied in the case of
modified gravity. Hence, we will restrict our discussion
to linear scales in this paper. 

\subsection{A cosmological probe of the Poisson equation}
\label{sec:Poisson}

Recently, techniques have been proposed to isolate the geometrical factors
in weak lensing correlations from the growth of structure, bias, and
matter power spectrum (shear ratios, or ``cosmography'' \cite{JainTaylor,ZhangHuiStebbins}). 
Because of the
separation of geometry from growth effects, cosmographic techniques apply 
equally well in modified gravity theories, i.e. when using the generalized
expressions~(\ref{eq:Cgg})--(\ref{eq:Ckk}).

A similar approach as for cosmography can be used 
to separate out the effect of a modified Poisson equation from the
growth factor and matter power spectrum. Observing a modification of the 
Poisson equation on cosmological scales would constitute a ``smoking gun'' 
for deviations from the GR + smooth Dark Energy scenario. 
Consider a very narrow galaxy redshift
distribution centered around $z_f$. Then, the galaxy-galaxy and
galaxy-shear correlation coefficients become:
\bea
C^{gg}(\l) &\simeq& \frac{H(z_f)}{\chi_f^2} b^2 
\left [ D_m^2(k,z)\:P(k) \right ]_{k=\frac{l+1/2}{\chi_f}}, \;\; \chi_f\equiv\chi(z_f)
\label{eq:Cgg-f} \\
C^{g\k}(\l) &\simeq& \frac{H(z_f)}{\chi_f^2} b W_\k(z_f)
\left [ \DPhim(k,z)\:D^2_m(k,z)\:k^2 P(k) \right ]_{k=\frac{l+1/2}{\chi_f}}
\label{eq:Cgk-f}
\eea
With this simplifying assumption, ratios of weak lensing correlations
can be used to isolate the effect of a modified Poisson equation:
\be
\P(\l) \equiv \frac{C^{g\k}(\l)}{C^{gg}(\l)} \simeq
b^{-1}\: H_0^2 W_\k(z_f)\:  \left ( \frac{k^2}{H_0^2}\DPhim(k,z_f)\right )_{k=\frac{\l+1/2}{\chi_f}}
\stackrel{{\rm GR+DE}}{=} \frac{3}{2}\Om \:b^{-1}\:H_0^2 W_\k(z_f)\:(1+z_f)
\label{eq:Poissonratio}
\ee
These ratios divide out the dependency on the matter power spectrum and
growth factor; they still depend on the expansion history, although not
very strongly, via $\chi_f$ and the lensing weight function $W_\k(z_f)$. 
On linear scales, $\P(\l)$ is independent
of scale in GR+DE, due to the scale-independence of both
$k^2\DPhim$ and bias $b$. In contrast, a scale-dependence is expected for
gravity theories which modify the Poisson equation. Hence, measuring $\P(\l)$ as
a function of scale $\l$ would constitute a robust test of the 
Poisson equation on large scales. See \refsec{DE} for a discussion of the
effect of non-smooth Dark Energy.

\section{Modified Gravity Models}
\label{sec:MG}

In this section, we briefly present the modified gravity models and
parameters adopted in this paper, and discuss the qualitative expectations
for their lensing predictions. First, let us discuss general constraints
applicable to any theory of gravity.
In any metric theory of gravity that conserves energy-momentum,
the evolution of super-horizon perturbations is essentially defined by the 
expansion history alone \cite{Bertschinger}. This can be understood
in the ``separate universes'' picture: a super-horizon curvature perturbation
behaves essentially as a separate Friedmann-Robertson-Walker Universe with
a constant small curvature. Hence, the evolution of super-horizon
curvature perturbations in modified gravity models is the same as in GR
(given the same expansion history) \cite{HuSawicki07,Bertschinger2}. 
Once a scale-independent super-horizon relation between 
the metric potentials $\Phi$ and $\Psi$ has been supplied by a theory,
the evolution of $\Phi$, $\Psi$ is fixed on super-horizon scales.
In the opposite limit, when a mode is well within the horizon, 
one can apply the quasi-static
approximation, neglecting time derivatives with respect to spatial
gradients. In this regime, the potentials $\Phi$, $\Psi$ are given by
a modified Poisson equation which again is determined by the theory.

Recently, Hu and Sawicki \cite{HuSawicki07} presented a parametrization of modified 
gravity theories which is based on the super-horizon and quasi-static
regimes, with a suitable interpolation between both. They showed that this
``Parametrized Post-Friedmann'' (PPF) approach
reproduces the predictions of the late-time acceleration
$\fR$ and DGP models well, with only few model-dependent functions.
Several alternative parametrizations of modified gravity have been 
proposed \cite{Amendola,Caldwell,JainZhang,Bertschinger2,SongKoyama}. 
The PPF parametrization is most useful for us in that it describes
the $\fR$ and DGP models with suitable parameters.

The two main model-dependent quantities with impact on
cosmological observables are (in the notation of \cite{HuSawicki07}),
(i) the metric ratio $g(k,z)$, given by:
\be
g(k,z) \equiv \left . \frac{\Phi+\Psi}{\Phi-\Psi}\right |_{k,z},
\label{eq:g}
\ee
and (ii) the rescaling $f_G(z)$ of the Newton constant in the Poisson equation
[\refeq{PoissonLCDM}] on sub-horizon scales:
\be
G_{\rm eff}(z) = \frac{G}{1+f_G(z)}.
\ee
Note that for GR coupled with a smooth Dark Energy, $g = f_G = 0$ at late times. 
Thus, $f_G$ influences $\DPhim$
directly, by rescaling the GR value by $1/(1+f_G)$. The metric ratio
$g$ influences the evolution of cosmological perturbations 
\cite{HuSawicki07,Bertschinger2}: if $g > 0$, i.e. $\Phi+\Psi>0$,
the potentials decay faster during the epoch of onsetting acceleration
than in the GR limit.
Conversely, for $g<0$ the potential decay is slowed down and, for
sufficiently negative $g$, inverted into a potential growth. From
\refeqs{Cgg}{Ckk}, we expect this suppressed or enhanced growth of
perturbations to lead to potentially observable effects in the weak
lensing correlations.

\begin{figure}[t]
\begin{center}
\includegraphics[width=.48\textwidth]{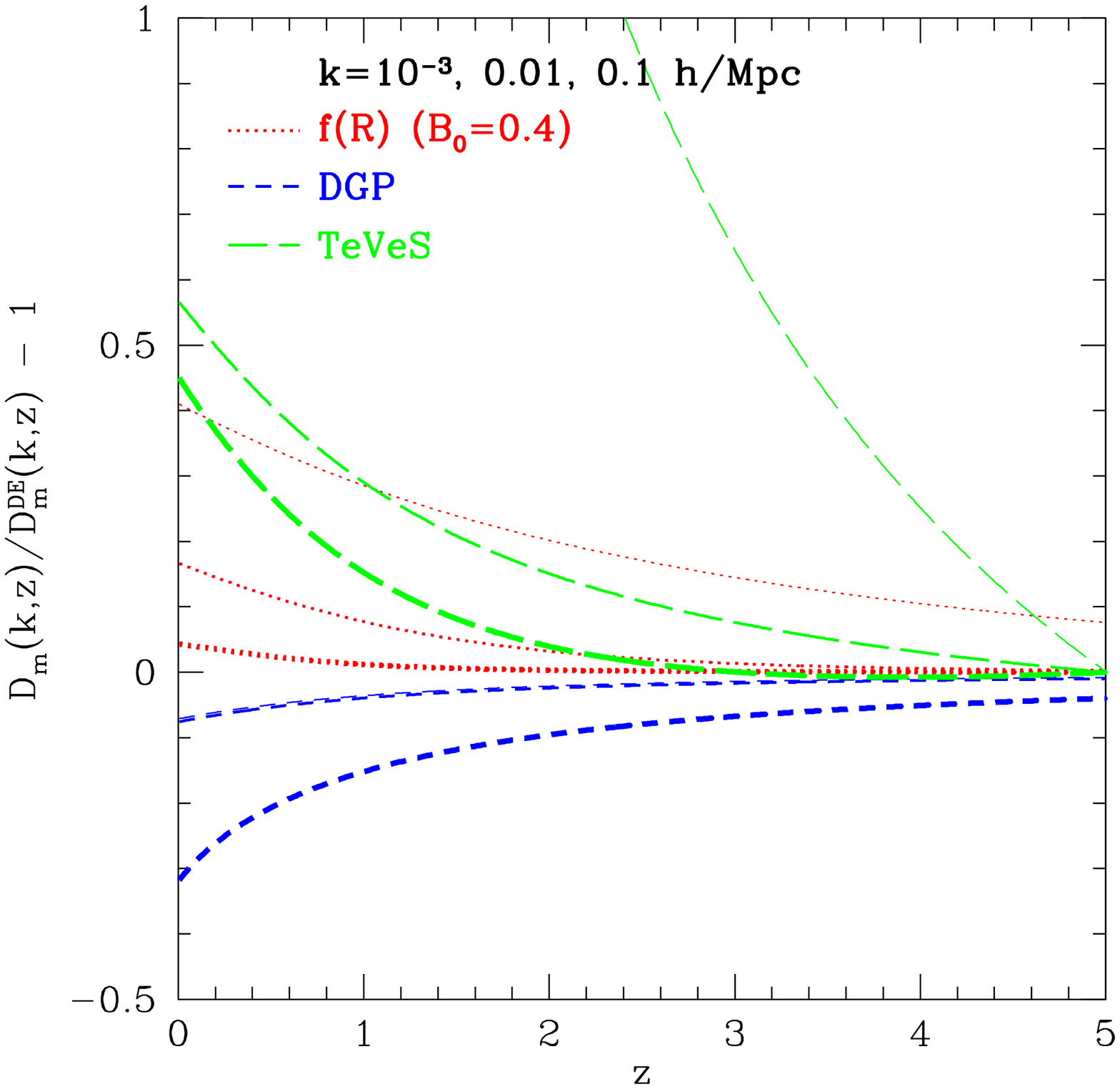}
\includegraphics[width=.48\textwidth]{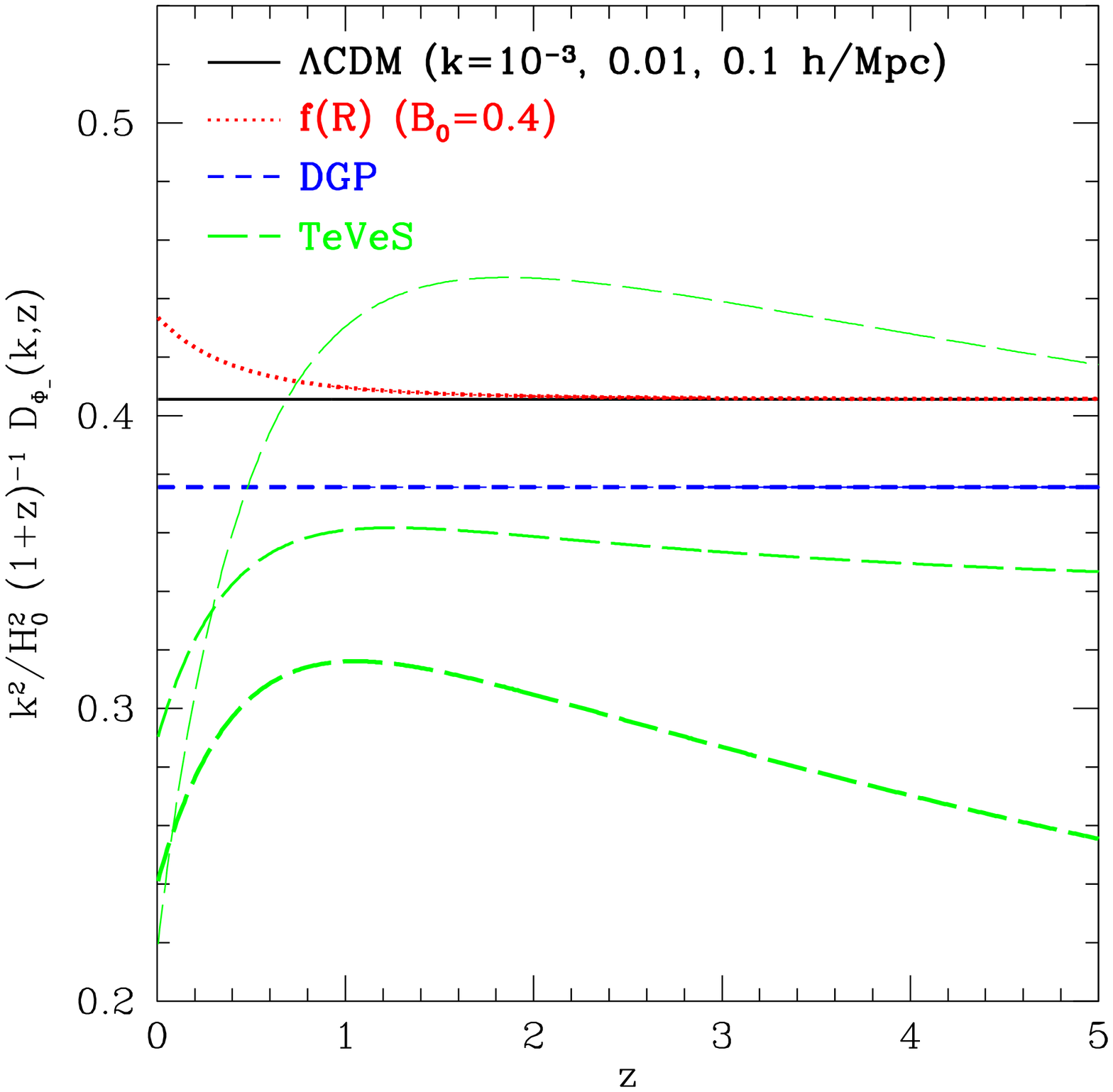}
\end{center}
\caption{\textit{Left panel:} Deviation of the growth factor $D_m(k,z)$
from a GR+DE model with the same expansion history, for three different 
modified gravity models: $f(R)$ (red/dotted; using a value of $B_0=0.4$),
DGP (blue/short-dashed), TeVeS (green/long-dashed; normalized to 1 at $z=5$ for
clarity). The three lines show different scales:
$k=10^{-3}\iMpch$ (thick), $k=0.01\iMpch$ (medium), and $k=0.1\iMpch$ (thin).
\textit{Right panel:} The Poisson factor $\DPhim$ [\refeq{Pratio}] scaled
by $k^2/H_0^2/(1+z)$, which reduces to $3\Om/2$ in the case of an unmodified
Poisson equation, as a function of redshift for $\Lambda$CDM and modified gravity 
models. The lines correspond to the same scales as in the left panel.
\label{fig:DPhim}}
\end{figure}

\subsection{$\fR$ model}

In $\fR$ gravity, the Einstein-Hilbert action is modified by adding
a certain function of the Ricci scalar, $f(R)$, to the gravitational part
of the Lagrangian \cite{CarrollEtal,Carroll2,Mota2006,NoOd2006,Faulkner06,BeanEtal07,SongEtal07,NoOd2008,Capozziello2008}:
\be
S_{{\rm EH}} = \frac{1}{16\pi\:G} \int d^4x \sqrt{-g}\:[ R + f(R) ]
\label{eq:fR-action}
\ee
In the case of $f(R) = C = \rm const.$, we recover $\Lambda$CDM, i.e. GR with
a cosmological constant $\Lambda = -C/2$. A linear function $f(R)$,
so that $f_R \equiv df/dR$ is a constant, simply amounts to a rescaling 
of Newton's constant, $G \rightarrow G/(1+f_R)$. Thus, 
a nontrivial modification of gravity
is linked to a nonzero second derivative of $f$.
The dimensionless field $f_R \equiv df/dR$ then appears as an additional
scalar degree of freedom in the modified Einstein equations.

Different functional forms of $\fR$ have been proposed in the literature. 
They generally satisfy the asymptotic relation 
$f(R)\rightarrow 0 $ for $R\rightarrow \infty$, so that GR is in principle restored
in the high-curvature limit, which applies to the Solar System as well as the
early Universe. Note that Solar System constraints strongly limit the
possible choices of $f$ \cite{Faulkner06,HuSawickifR,NoOd2008}. In addition, 
$d^2f/dR^2$ should be positive in order to
achieve a stable high-curvature limit \cite{BeanEtal07,SongEtal07}. 
For this paper,
we adopt the implicit definition of $\fR$ proposed in \cite{SongEtal07}.
Since the background curvature $R$ is fixed as a function of $a$ for
a given expansion history $a(t)$, the function $f(R(a))$ can be
determined from the modified Friedmann equations, which are obtained by applying
the modified Einstein equations to the FRW metric. We assume the
expansion history of our fiducial $\Lambda$CDM model, defined by $\Om=0.27$, 
$\OL=0.73$, $\Omega_b=0.046$, $h=0.7$, $n_s=0.95$. For the power spectrum
normalization, we use the CMB normalization of the primordial curvature
perturbations given by 
$\d_\zeta=4.58\cdot 10^{-5}$ at $k=0.05\: \rm Mpc^{-1}$, as in the case 
of the $\Lambda$CDM model. This is motivated by the fact that this $\fR$ model
makes definite predictions for the CMB (i.e., identical to $\Lambda$CDM), 
whereas measurements of the power spectrum amplitude today, e.g. using
the power spectrum of galaxies, are affected by
gravitational non-linearities, a regime where definite predictions for modified
gravity are still outstanding.

The amplitude of
the field $f_R$, and correspondingly of the modification to gravity, 
is most conveniently parametrized in terms of the parameter 
$B_0=B(a=1)$, where $B(a)$ is a function proportional to the second derivative
of $f$:
\be
B(a) \equiv \frac{d^2f/dR^2}{1+f_R}R' \frac{H}{H'},
\label{eq:Bdef}
\ee 
where $R(a)$ is the Ricci scalar of the FRW background, which in turn 
can be expressed in terms of the Hubble rate $H(a)$ and its derivatives,
and primes denote derivatives with respect to $\ln a$. Thus, the evolution
of $f(R)$ in the background is fixed by the expansion history together with $B_0$,
and we then proceed to calculate the evolution of
matter perturbations and potentials using the PPF parametrization
presented in \cite{HuSawicki07}.

For a positive value of $B$, necessary for stability in the high-curvature 
limit, $f(R)$ models generically predict a \textit{negative} $\Phi+\Psi$, i.e.,
a metric ratio $g < 0$. On super-horizon scales, $g$ is of order $-B$, and
grows to a value of $-1/3$ on small scales. The negative potential ratio 
leads to a slower decay of the potentials
during the acceleration epoch, and hence a smaller growth suppression
of matter perturbations (\reffig{DPhim}, left panel) with respect to
$\Lambda$CDM. Having fixed the fluctuation amplitude at early times through 
the CMB, we thus expect an increased weak lensing signal from the $\fR$ model.

As pointed out above, the gravitational constant is rescaled by
$1/(1+f_R)$ in $\fR$ models. Since $f(R) < 0$ to achieve
acceleration, $G_{\rm eff} > G$ in $\fR$ gravity, and hence $\DPhim$ is
larger than its $\Lambda$CDM value (right panel of \reffig{DPhim}). However, this 
effect is quite small, at the percent level, for observationally allowed models.
For the weak lensing forecasts below, we adopt a value of $B_0=0.1$,
leading to a field amplitude today of $f_R(a=1) = -0.017$. For purposes
of presentation,
we use the larger value of $B_0=0.4$ for the plots in \reffig{DPhim}.

\subsection{DGP model}
\label{sec:DGP}

In the Dvali-Gabadadze-Porrati model \cite{DGP,Deffayet01,DeffayetEtal02}, 
all matter
and non-gravitational interactions are confined to a (3+1)-dimensional
``brane'' embedded in (4+1)-dimensional space. Only gravity is 
five-dimensional, and the effects of the large extra dimension become
noticeable at the crossover scale $r_c = G^{(5)}/2G^{(4)}$,
where $G^{(5)}$ is the gravitational constant in five dimensions,
and $G^{(4)}$ is the usual four-dimensional gravitational constant.
The crossover scale $r_c$ is the only free parameter of the model.
One of the two possible branches of the model
naturally leads to an accelerated expansion when the 
horizon scale $c/H(z)$ reaches $r_c$.
The expansion history in the DGP model is somewhat different than
$\Lambda$CDM. It is given by \cite{Deffayet01}:
\be
H(z) = H_0 \left ( \sqrt{\Omega_{rc}}+\sqrt{\Omega_{rc}+\Om a^{-3}} \right ),
\ee
where $\Omega_{rc} = 1/(4r_c^2H_0^2)$, and, for a flat Universe,
$\Omega_{rc} = (1-\Om)^2/4$. Here, we adopt $\Om=0.25$ and $h=0.66$, as 
determined from fitting Supernova distances and the CMB \cite{SongEtalDGP},
corresponding to $r_c H_0 \approx 1.78$.
For the same reasons as in the case of the $\fR$ model, we assume a primordial 
power spectrum normalization of 
$\d_\zeta=4.58\cdot 10^{-5}$ at $k=0.05\:\rm Mpc^{-1}$ for the DGP model.
For the evolution of perturbations in DGP, effects from the 5D nature of
gravity, namely perturbations in the extrinsic curvature of the brane,
have to be taken into account in order to close the perturbation 
equations \cite{LueEtal04,KoyamaMaartens}. Here, we again employ the 
interpolation based on the PPF parametrization given in \cite{HuSawicki07}.
In order to compare the DGP predictions with those from GR+DE, we use a 
Dark Energy model
with an equation of state $w_{\rm eff}(a)$ which exactly mimics the
expansion history of the DGP model.

The DGP model has a positive metric ratio $g$ which grows proportional
to $1/(H\,r_c)$ and is of order unity at $z=0$; it is only weakly 
scale-dependent for sub-horizon scales. The positive $g$ leads to an
enhanced potential decay with respect to GR+DE \cite{SongEtalDGP}, as is
apparent in \reffig{DPhim} (left panel).
Correspondingly, we expect a reduced lensing signal for the DGP model. 
Note that the decay of the
potentials sets in at considerably higher redshifts, $z\sim 5$, than in GR+DE.
This is because $w_{\rm eff} \approx -1/2$ in the DGP model for $z \gtrsim 1$
\cite{DvaliTurner}.
The Poisson equation is not modified in the DGP model. However,
as the Poisson ratio $\P(\l)$ is proportional to $\Om$ [\refeq{Poissonratio}],
the prediction for the DGP model shows a scale-independent departure due to 
the slightly smaller value of $\Om$ adopted for this model  (right panel of \reffig{DPhim}).
In contrast to the simple DGP model considered here, generalized 
braneworld-inspired modified gravity models can show a modified Poisson 
equation \cite{Koyama2006}.

\subsection{TeVeS}

The TeVeS model \cite{Bekenstein} is a relativistic metric theory of gravity 
which, as the models presented above, reduces to General Relativity in the 
high-density, high acceleration regime. However, in the high-density,
weak acceleration regime, it behaves like MOdified Newtonian Dynamics
(MOND, \cite{MOND}). The MOND force law can explain galactic rotation 
curves without invoking dark matter. \textit{TeVeS} generalizes this idea by
adding an additional vector field (\textit{Ve}) and scalar field (\textit{S})
in such a way that the same effect of dark matter apparently seen in
dynamics is also seen in gravitational lensing. Hence, TeVeS can be
taken as an attempt to explain all cosmological observables without 
any dark matter of unknown nature.

For this paper, we adopt the ``neutrino model'' from \cite{Schmidt07}.
The matter content in this model is given by baryons, $\Omega_b = 0.05$, 
and neutrinos, $\Omega_\nu = 0.17$, with a cosmological constant given by
$\Omega_\Lambda = 0.78$, in order to match the observed expansion
history. This model has been shown to be in acceptable
agreement with CMB and matter power spectrum observations \cite{Skordis1}.
While the current TeVeS model uses a cosmological constant to achieve
acceleration, related theories invoking vector fields have been
proposed which lead to acceleration without a cosmological constant
\cite{Zhao}.
The evolution equations of linear cosmological perturbations in TeVeS 
have been derived in \cite{Skordis1,Skordis2,Dodelson,Bourliot}. In order
to satisfy Big Bang nucleosynthesis bounds, the TeVeS scalar field is 
constrained to small values throughout cosmological history, and neither
significantly contributes to the background expansion nor the growth
of perturbations. In contrast,
perturbations of the time-like vector field exhibit a growing mode which in 
turn boosts the growth of matter perturbations \cite{Dodelson}. Hence,
the perturbations of the vector field in TeVeS play the role of seeds
of structure formation in the absence of dark matter.
The vector degree of freedom is formed by the spatial components of the time-like
vector field, which couple to the spatial part of the metric. Hence,
the growing vector field only affects the potential $\Phi$, which
leads to a nonzero $\Phi+\Psi < 0$, i.e. $g < 0$ \cite{Dodelson}. 
As in the case of the $\fR$ model, this produces a slower decay, or
even growth, of the potentials at late times. However, the power spectrum
at earlier times is lower than that of $\Lambda$CDM due to the considerably
different history of structure formation in TeVeS.

In addition to their effect on the potential ratio, 
the vector perturbations also appear
on the right-hand side of the Poisson equation \cite{Schmidt07}, and even
dominate the matter perturbations on small scales. This effect can be
seen in the ``Poisson factor'' (\reffig{DPhim}, right panel): on large
scales (thick line), the value is significantly smaller than in $\Lambda$CDM
due to the smaller $\Om$ in TeVeS without dark matter. Going to smaller scales,
$\DPhim$ grows and eventually overtakes the $\Lambda$CDM value due to the increasing
contribution of vector perturbations \cite{Schmidt07}. 
Thus, the Poisson factor in TeVeS shows a 
characteristic scale dependence and redshift evolution, which should be
observable in the correlation ratio $\P(\l)$ defined in \refsec{Poisson}.
We return to this in \refsec{Pratio}.

The TeVeS modifications to GR are somewhat more severe than those from
the late-time accelerating models $\fR$ and DGP; a simple parametrization
in terms of $g(k,z)$ and $f_G(z)$ does not exist (yet) for this model.
Since the TeVeS model used here has a slightly different expansion history 
than $\Lambda$CDM,
we compare it with an effective GR+DE model with the same expansion history.

\subsection{Comparison with general Dark Energy models}
\label{sec:DE}

In the previous sections we pointed out the two main physical quantities
affected by gravity, aside from the expansion history: the growth factor
of matter perturbations, $D_m(k,z)$, and the matter-potential relation,
$\DPhim(k,z)$. In modified gravity, they are influenced by a non-zero
metric ratio $g$, and a rescaling of the gravitational constant or additional
degrees of freedom appearing in the Poisson equation.
Can these signatures possibly be mimicked by a general
Dark Energy model ? 

Within GR, anisotropic stress perturbations are the only source of differing
cosmological potentials. In order to achieve a given metric ratio $g(k,z)$, so that
$\Phi+\Psi = g\:\Phi_-$, one would need components of the dimensionless 
anisotropic stress perturbation $\Pi$, at late times, 
of the order of \cite{Bertschinger2}:
\be
\Pi(k,z) \sim \frac{a\:k^2}{H_0^2} g(k,z) \Phi_-(k,z). 
\ee
Clearly, the components of the Dark Energy anisotropic stress tensor have
to be large on small scales in order to produce the order unity values of $g$
predicted by modified gravity theories. Any neutrino contribution to $\Pi$
will be very small at late times. In addition, Dark Energy density perturbations
would also add a contribution of the order of $\rho_{DE}\d_{DE}/(\rho_m\d)$
to the Poisson ratio $\P(\l)$ introduced in 
\refsec{Poisson}. While the Dark Energy properties necessary to emulate 
modified gravity seem not to be very natural ones, it is not possible 
to place stringent constraints on general Dark Energy models, in lack of an 
underlying theory. In the linear regime, the two functions $\d_{DE}(k,z)$ and
$\Pi(k,z)$ are sufficient to emulate any given modified gravity model. 
However, one might expect that more freedom is needed for a Dark Energy model
to extend this emulation into the non-linear regime. For example,
the $\fR$ and DGP models exhibit a chameleon-like behavior \cite{Khoury04}
in order to restore GR in high-density environments 
\cite{NavarroAcoleyen,Faulkner06,DeffayetEtalDGPNL}. To emulate this effect,
a Dark Energy model would also need a chameleon-like coupling to matter
\cite{Khoury04}. It would be worth studying to what extent general physical 
constraints can be placed on Dark Energy models which also apply in 
the non-linear regime.

\begin{figure}[t]
\begin{center}
\includegraphics[width=.48\textwidth]{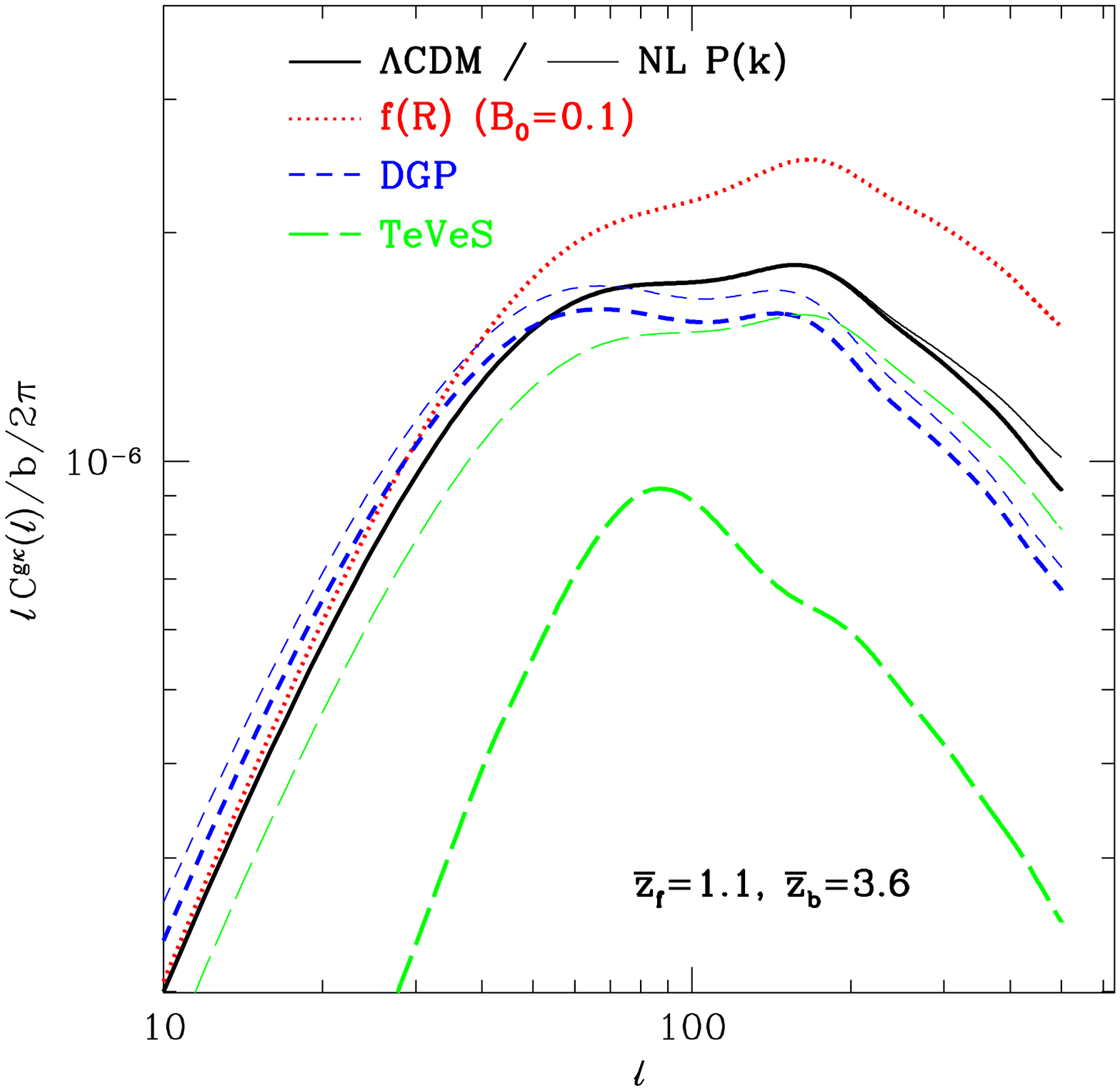}
\includegraphics[width=.48\textwidth]{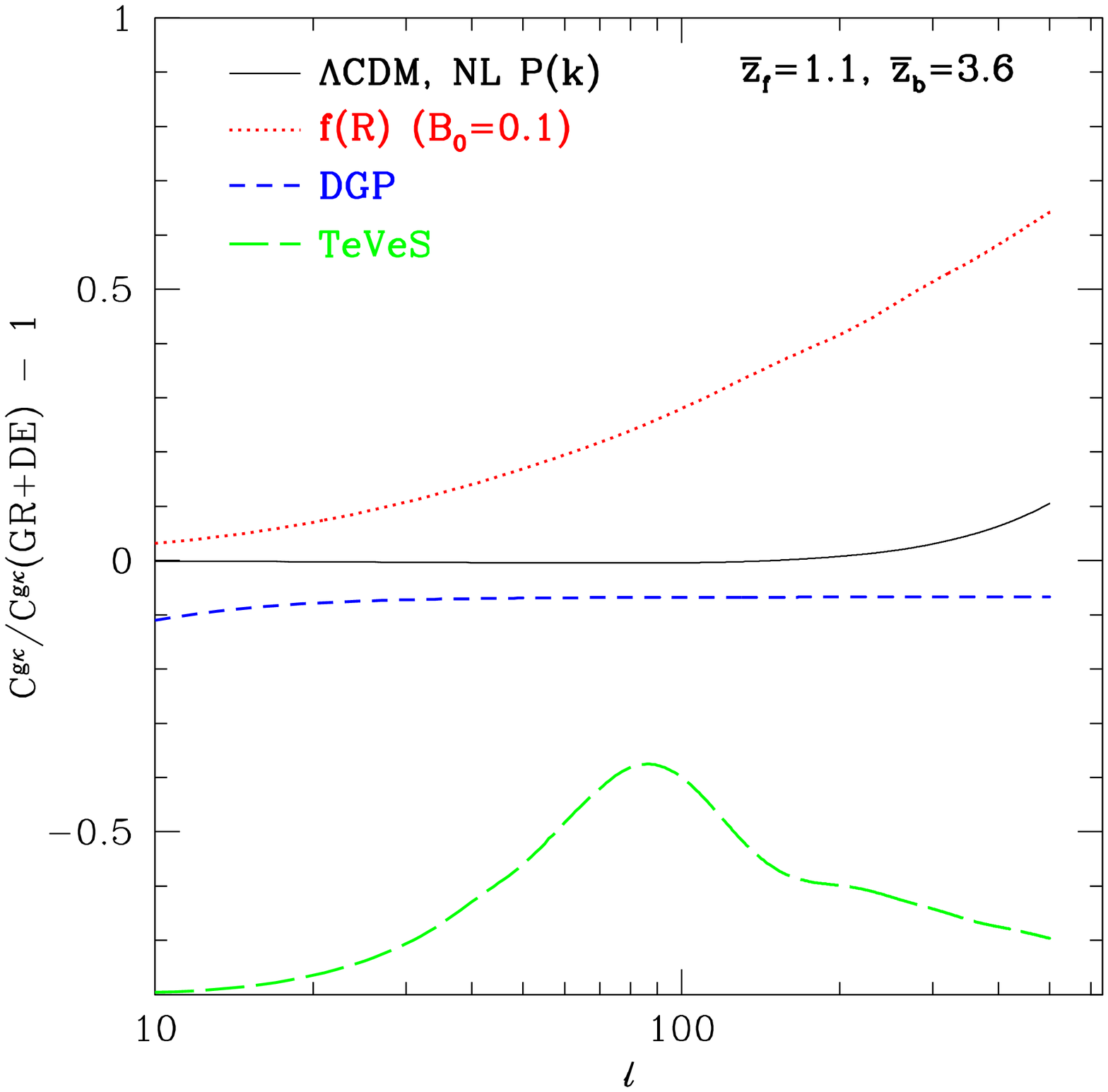}
\end{center}
\caption{\textit{Left panel:} The galaxy-shear cross power $C^{g\k}(\l)/b$
for $\Lambda$CDM (black/solid) and modified gravity theories: $f(R)$ (red/dotted),
DGP (blue/short-dashed), and TeVeS (green/long-dashed). In case of DGP and
TeVeS, the thin lines show the corresponding predictions for a GR+DE model 
with the same expansion history.
The foreground galaxies are from bin ``F'' with
median redshift of $\zbar_f = 1.1$, and background (sheared) galaxies 
are from bin ``B'' with $\zbar_b=3.6$ (see text).
\textit{Right panel:} Relative deviation of the galaxy-shear cross power 
$C^{g\k}(\l)$ of modified gravity models from that of a GR+DE model with 
identical expansion history, for the same redshift bins as in the left panel.
\label{fig:Cgk-vs-l}}
\end{figure}

\begin{figure}[t]
\begin{center}
\includegraphics[width=.48\textwidth]{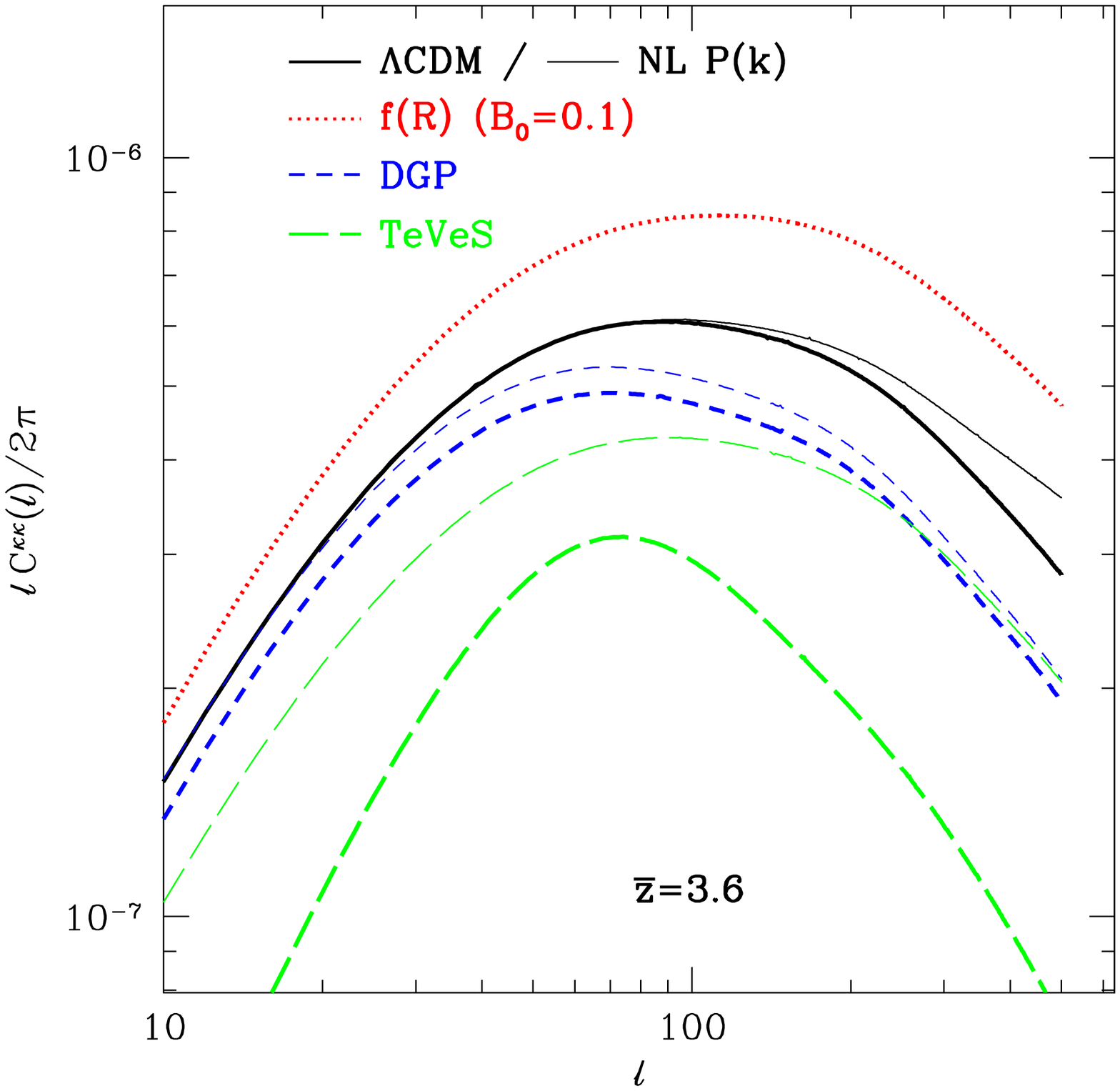}
\includegraphics[width=.48\textwidth]{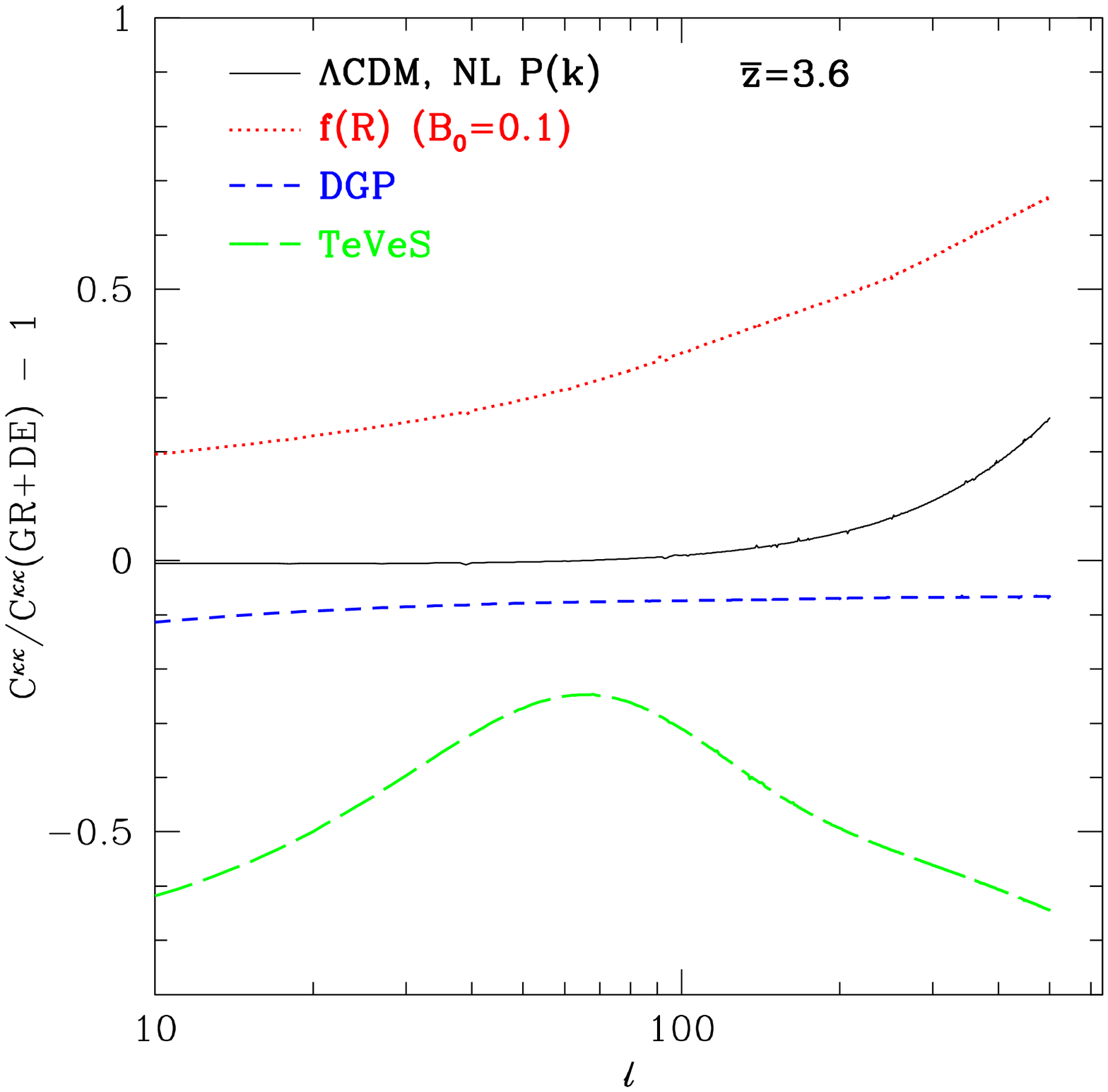}
\end{center}
\caption{\textit{Left panel:} The shear-shear auto-correlation $C^{\k\k}(\l)$
for $\Lambda$CDM and modified gravity theories, for high-redshift galaxies with
median redshift $\zbar_b=3.6$. The different colors and line styles correspond
to the same modified gravity and GR+DE models as in \reffig{Cgk-vs-l}.
\textit{Right panel:} Relative deviation of the shear power spectrum $C^{\k\k}(\l)$
from that of GR+DE models with the same expansion history, for the same 
high-redshift galaxies as in the left panel.
\label{fig:Ckk-vs-l}}
\end{figure}

\begin{figure}[t]
\begin{center}
\includegraphics[width=.48\textwidth]{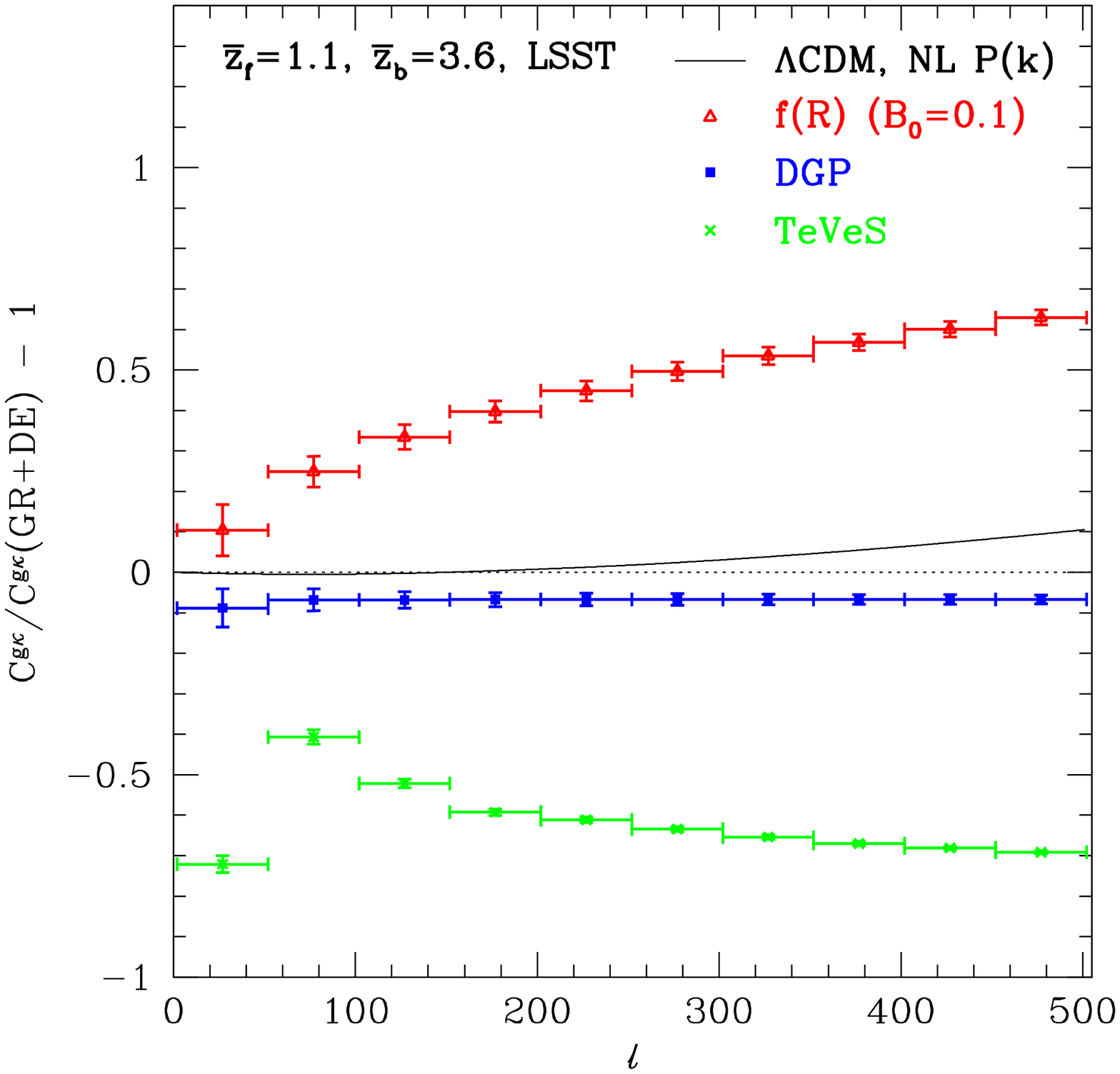}
\includegraphics[width=.48\textwidth]{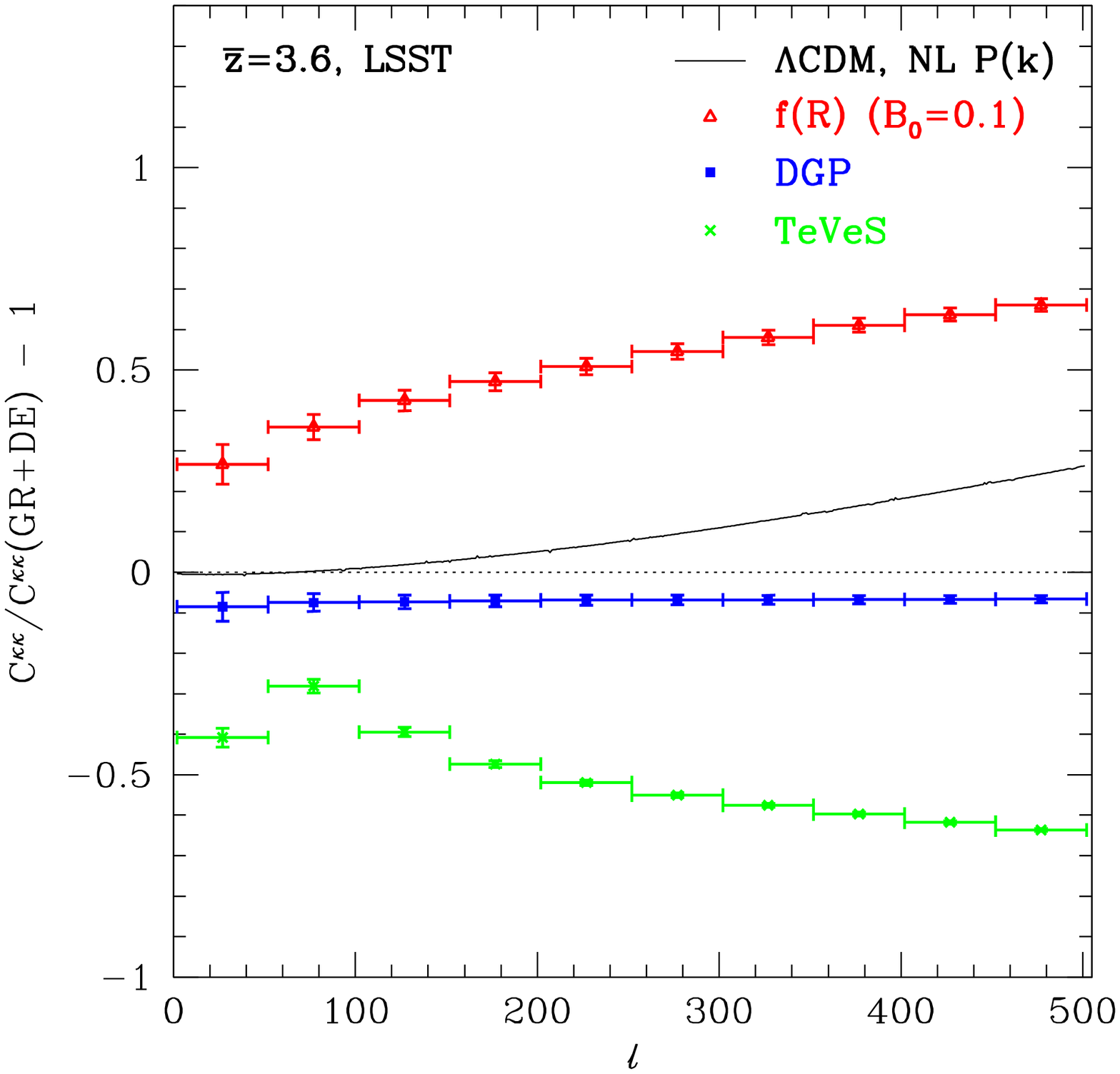}
\end{center}
\caption{\textit{Left panel:} Relative deviation of the cross power 
$C^{g\k}(\l)$ from that of the corresponding GR+DE models, for the same redshift 
bins as in \reffig{Cgk-vs-l}, but in bins of
$\Delta\l = 50$ with statistical errors expected from LSST.
\textit{Right panel:} Same as the left panel, but for the
shear-shear correlation of high-$z$ galaxies (as in \reffig{Ckk-vs-l}).
\label{fig:C-vs-l-werr}}
\end{figure}

\section{Forecasts for weak lensing surveys}
\label{sec:res}

We now present quantitative forecasts of the constraining power of
future surveys with regard to modified gravity. Following the treatment
presented in \refsec{WL}, we consider the
galaxy-shear and shear-shear correlations, $C^{g\k}$, and $C^{\k\k}$, on
linear scales ($\l \lesssim 200$ for $z \gtrsim 1$). In this regime, 
the Gaussian
error estimate is a good approximation. The variance of the
cross-correlation $C^{AB}(\l)$, where $A,B$ stand for any of $g_i,\k_j$, is 
then given by:
\be
[\Delta C^{AB}(\l)]^2 = \frac{1}{\fSky (2\l+1)} \left \{
\left [ C^{AA}(\l) + \frac{\sigma_A^2}{n_A} \right ] \left [ C^{BB}(\l) + \frac{\sigma_B^2}{n_B} \right ]
+ C^{AB}(\l)^2 \right \}.
\label{eq:DeltaCl}
\ee
Here, $\fSky$ is the fraction of sky observed in the survey, $n_{A,B}$
denote the number densities of galaxies per sr in each sample, and 
$\sigma_{A,B}$ is the shot noise error. For galaxy number counts,
$\sigma_g=1$, while for the shear (convergence), we take $\sigma_\epsilon=0.35$ as
the shot noise due to intrinsic ellipticities. We use the exact integrals
[\refeqs{CggEx}{CkkEx}] for the correlations for $\l \leq 10$. For higher $\l$,
we use the Limber-approximation expressions [\refeqs{Cgg}{Ckk}], since the 
deviations are less than 1\% and drop quickly for higher $\l$.
Note that in any case the bulk of the signal-to-noise comes from much
 higher $\l$.

In principle one should take into account magnification bias
as well, which adds additional fluctuations to the observed galaxy
overdensity and hence affects $C^{gg}$ and $C^{g\k}$. 
For a galaxy population with bias $b$, the observed
galaxy overdensity $\d_g$ will be given by \cite{LoVerdeHuiGaztanaga2007}:
\be
\d_g = b\:\d + (5s-2) \k,
\ee
where $\d$ is the matter overdensity, $\k$ is the convergence, and 
$s = d\lg N/dm$ is the logarithmic number count slope as a function of
magnitude $m$. Typical values for galaxy surveys are $s=0.2\dots 0.6$
\cite{LoVerdeHuiGaztanaga2007,HuiGaztanagaLoVerde2007}. The galaxy-shear
correlation will thus be modified as:
\be
C^{g\k}(\l) \rightarrow C^{g\k}(\l) + (5s-2)\: C^{\k_f\k}(\l),
\ee
where $\k_f$ is the convergence calculated for the foreground galaxy redshift
bin. Note that modified gravity will also affect the magnification bias term.
Since we do not have a realistic estimate
for the number count slope in the surveys considered here, we do not
include this effect. For the galaxy-shear correlation with foreground galaxies
centered around $z=1.1$ and for the range in $s$ given above, we estimate
that the effect is around 10\% ($\l\gtrsim 50$) for $\Lambda$CDM and reaches
up to 30\% for the $f(R)$ cosmology. Note that the parameter $s$ is measurable
in surveys. Furthermore, by varying the magnitude cut of the galaxy sample,
one can vary $s$ to some extent and in this way probe the magnification effect.
Hence, while straightforward to include for an actual survey, 
magnification bias will not affect the forecasts presented here appreciably.

Our fiducial flat $\Lambda$CDM cosmology is given by: $\sigma_8=0.8$, $\Om=0.27$, 
$\OL=0.73$, $\Omega_b=0.046$, $h=0.7$, $n_s=0.95$. For the linear matter
power spectrum, we use the transfer function from \cite{EisensteinHu}.
The parameters used for the modified gravity models are stated in 
\refsec{MG}. Additionally, we will consider a $\Lambda$CDM cosmology with a non-linear
power spectrum from \texttt{halofit} \cite{halofit}, in order to assess where
the linear approximation breaks down. Note that this is only indicative;
non-linearities can in principle become important already at larger scales in
modified gravity theories, due to the non-linear nature of the modified
evolution equations for perturbations.

\subsection{Survey specifications}
\label{sec:survey}

We adopt simple survey specifications, as we are mainly interested in 
the differences between modified gravity models from GR in this paper, which
do not depend appreciably on the detailed specifics of the survey. 
The galaxy redshift distribution $dN/dz$ is adopted
from observations \cite{Gabash2004,Gabash2006} for a magnitude-limited sample with 
$I < 27$, as parametrized as ``Sample I'' in \cite{LoVerdeISW}. This
is roughly what is expected to be attained by LSST \cite{LSST} or SNAP
\cite{SNAP}. The median
redshift of the distribution is 0.91. We then divide the galaxy distribution
into redshift bins defined by $z_l$, $z_h$:
\be
W_g(z) = C \frac{dN}{dz} \left [ \erfc\left ( \frac{z_l-z}{\sqrt{2}\sigma_z} \right )
- \erfc\left ( \frac{z_h - z}{\sqrt{2}\sigma_z} \right ) \right ],
\label{eq:z-bin}
\ee
where $C$ is a normalization constant, $\erfc$ is the complementary error
function, $z_l$ and $z_h$ are the lower 
and upper bin boundaries, respectively, and $\sigma_z$ is the expected 
photometric redshift error. We adopt $\sigma_z=0.03(1+z)$.

In the following, for studies of the \textit{redshift evolution} of the 
galaxy-shear and shear-shear correlations, we divide the 
redshift range from $z=0-3$ in bins
with $\Delta z = z_h-z_l = 0.4$, so that the first bin is defined by
$z_l=0; z_h=0.4$, while the last bin corresponds to $z_l=2.4; z_h=2.8$.
In addition, we define a ``background'' high-$z$ bin
encompassing the galaxies from $z=3$ to $z=5$ (bin ``B'', with a median 
redshift $\bar{z}_b=3.6$). We then correlate foreground galaxies or shear
with the shear of galaxies in bin ``B''.
For studies of the \textit{scale-dependence} of $C^{g\k}(\l)$ and $C^{\k\k}(\l)$, 
we choose the same background galaxies, and use a wider foreground redshift
bin defined by $z_l=0.8$, $z_h=1.6$ (yielding a median redshift of $z_f=1.1$).
The constraints on modified gravity do not depend strongly on the
number and precise redshift of the foreground and background bins 
chosen.

We do not attempt to model the galaxy bias in each bin, instead we show 
correlations
divided by $b$, where applicable. In practice, the bias can be taken as free 
(scale-independent) parameter to be marginalized ober. We will see that the
dependence on scale and redshift of the effects of modified gravity should
allow them to be disentangled from galaxy bias. 
We will show
forecasts for LSST, adopting $n_g=50$~arcmin$^{-2}$ as the total observed
galaxy density, and $\fSky=0.5$ as the observed sky fraction. 
Constraints similar to those for LSST are expected from the ``wide survey'' 
of SNAP \cite{SNAP}.
Galaxy-shear and shear-shear correlations offer a large amount of information.
As we are mainly interested in the specific behavior of weak lensing 
correlations in the different modified gravity models, we do not attempt to 
fully exploit this information here; instead,
we will separately discuss the scale ($\l$-)dependence and redshift evolution
of the observables, pointing out generic features of the different model
predictions, and whether these effects will be observable in future surveys.


\subsection{Scale dependence}

We first consider the galaxy-shear cross-correlation, correlating
foreground galaxies from bin ``F'' ($\bar{z}_f=1.1$) with the shear 
from galaxies in bin ``B'' ($\bar{z}_b=3.6$). \reffig{Cgk-vs-l} (left panel)
shows $C^{g\k}(\l)/b$ for $\Lambda$CDM and the modified gravity models,
where $b$ is the bias of the foreground galaxies.
There are clear differences in the magnitude of the signal, with the
largest correlation coming from the $\fR$ model, while DGP and TeVeS
yield weaker correlations than the corresponding GR+DE models with identical 
$H(z)$ (thin lines). 
Qualitatively, this is what we 
expected after the discussion of the models in \refsec{MG} (see also
\reffig{DPhim}). The differences to the predictions of the GR+DE models
with the same expansion history are of order
50\% or more in the case of $\fR$ and TeVeS, and of order 
10\% in the case of DGP
(right panel of \reffig{Cgk-vs-l}). 
A scale-dependence of the ratio is also apparent; in the case of the DGP model,
the deviations become slightly smaller towards smaller scales, while
the opposite scale dependence is apparent for the $\fR$ model.
This is due to the different values of the metric ratio in the
quasi-static (small scale) regime for the different models (\refsec{MG}).
Note that for DGP and $\fR$, we expect the GR behavior to be restored on 
sufficiently small (non-linear) scales. In the case of
TeVeS, the differences to GR are also sourced by the modified Poisson
equation, and hence remain present even at small scales.
Apparently, for the redshift bins chosen, non-linear evolution becomes 
relevant for $\l \gtrsim 300$ in the $\Lambda$CDM case. Hence, any predictions
at larger $\l$ should not be taken at face value. We found that the
deviations from GR in terms of $R^{g\k}(\l)=C^{g\k}(\l)/\sqrt{C^{gg}(\l)}$,
which is independent of the linear galaxy bias, are very similar to those of
$C^{g\k}$.

In \reffig{Ckk-vs-l} we show the corresponding results for the 
shear-shear correlation of the high-$z$ ``B'' bin.
The results are quite
similar to those for the galaxy-shear correlation. 
The shear-shear correlation offers the advantage of being independent of
any galaxy bias. However,
small-scale modes at low redshifts contribute to the 
shear-shear correlation, so that the effects of non-linear evolution become
relevant already at $\l\sim 200$.
In contrast, the contribution to the galaxy-shear correlation we considered
is concentrated around $\zbar_f=1.1$, where the non-linear evolution is 
somewhat less relevant.

Finally, we determine whether the differences in magnitude and 
scale-dependence of weak lensing correlations will be observable in
future surveys. \reffig{C-vs-l-werr} shows the deviations for
$C^{g\k}(\l)$ and $C^{\k\k}(\l)$ for the same redshift bins as above,
binned in $\l$ with $\Delta\l=50$. Also shown are $1\sigma$ statistical errors 
expected for an LSST-like survey, using \refeq{DeltaCl}. Apparently, the
overall difference in the lensing signal is distinguishable at very
high significance, to a comparable degree in both galaxy-shear and
shear-shear correlations. As both observables are
prone to different experimental systematics, the possibility of measuring 
both correlations
in the same survey will enable powerful cross-checks of any signs of deviations
from the GR predictions.
Choosing a more conservative background redshift bin from $z=2$ to 3, 
with median redshift
$\bar{z}_b=2.4$, degrades the expected signal-to-noise of the deviations
in $C^{g\k}$ and $C^{\k\k}$ by not more than 20\%.

In addition to the overall magnitude, the different scale dependence of 
$C^{g\k}(\l)$ and $C^{\k\k}(\l)$ in the $\fR$ and TeVeS models should
be clearly distinguishable, while the scale dependence of the deviation
of the DGP model might be difficult to detect.
A scale dependence of the deviation of lensing correlations from the
GR+DE prediction should break degeneracies with other cosmological 
parameters such as $b$ and $\sigma_8$. The redshift evolution of the lensing
signal, to which we now turn, can serve as an additional tool for this
purpose.

\begin{figure}[t]
\begin{center}
\includegraphics[width=.48\textwidth]{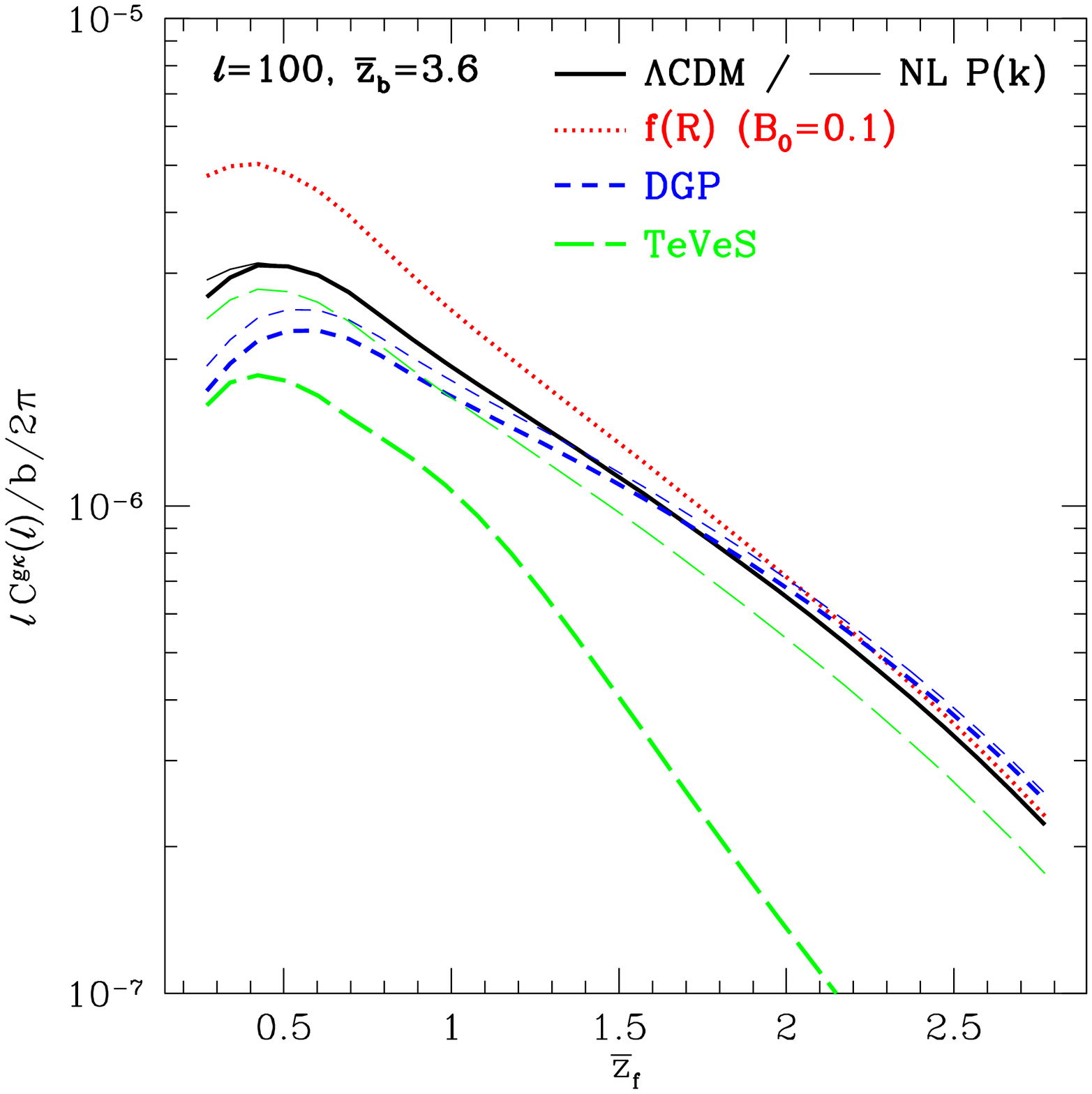}
\includegraphics[width=.48\textwidth]{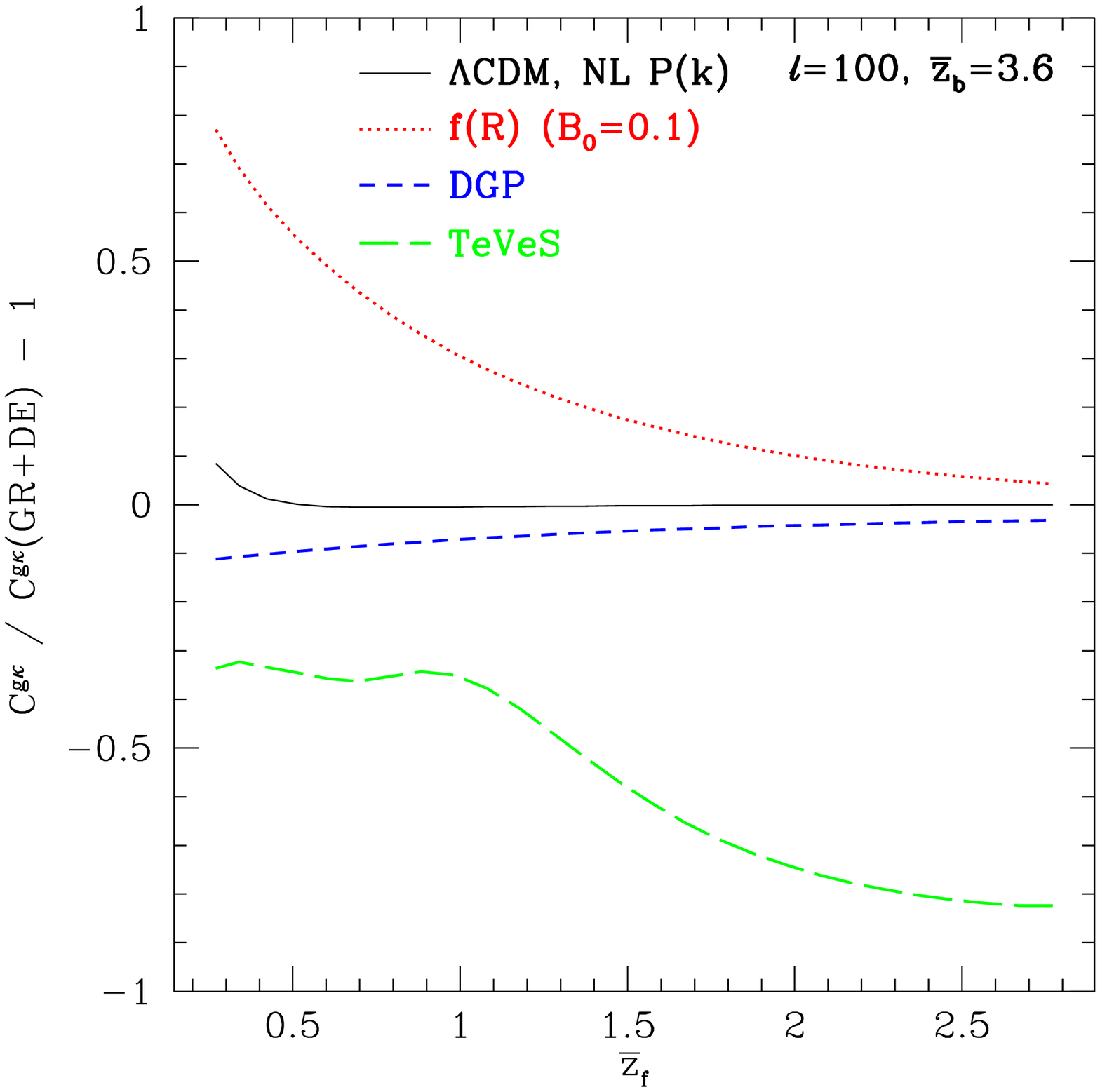}
\end{center}
\caption{\textit{Left panel:} The galaxy-shear 
cross power $C^{g\k}(\l)/b$ at $\l=100$ for a fixed 
background galaxy sample with median redshift $\zbar_b=3.6$, as 
a function of the redshift $\zbar_f$ of foreground galaxies. The different
lines correspond to the same modified gravity and GR+DE models as in
\reffig{Cgk-vs-l}.
\textit{Right panel:} Relative deviation of the galaxy-shear power
predicted by the modified gravity models from those of the corresponding
GR+DE models.
\label{fig:Cgk-vs-z}}
\end{figure}

\begin{figure}[t]
\begin{center}
\includegraphics[width=.48\textwidth]{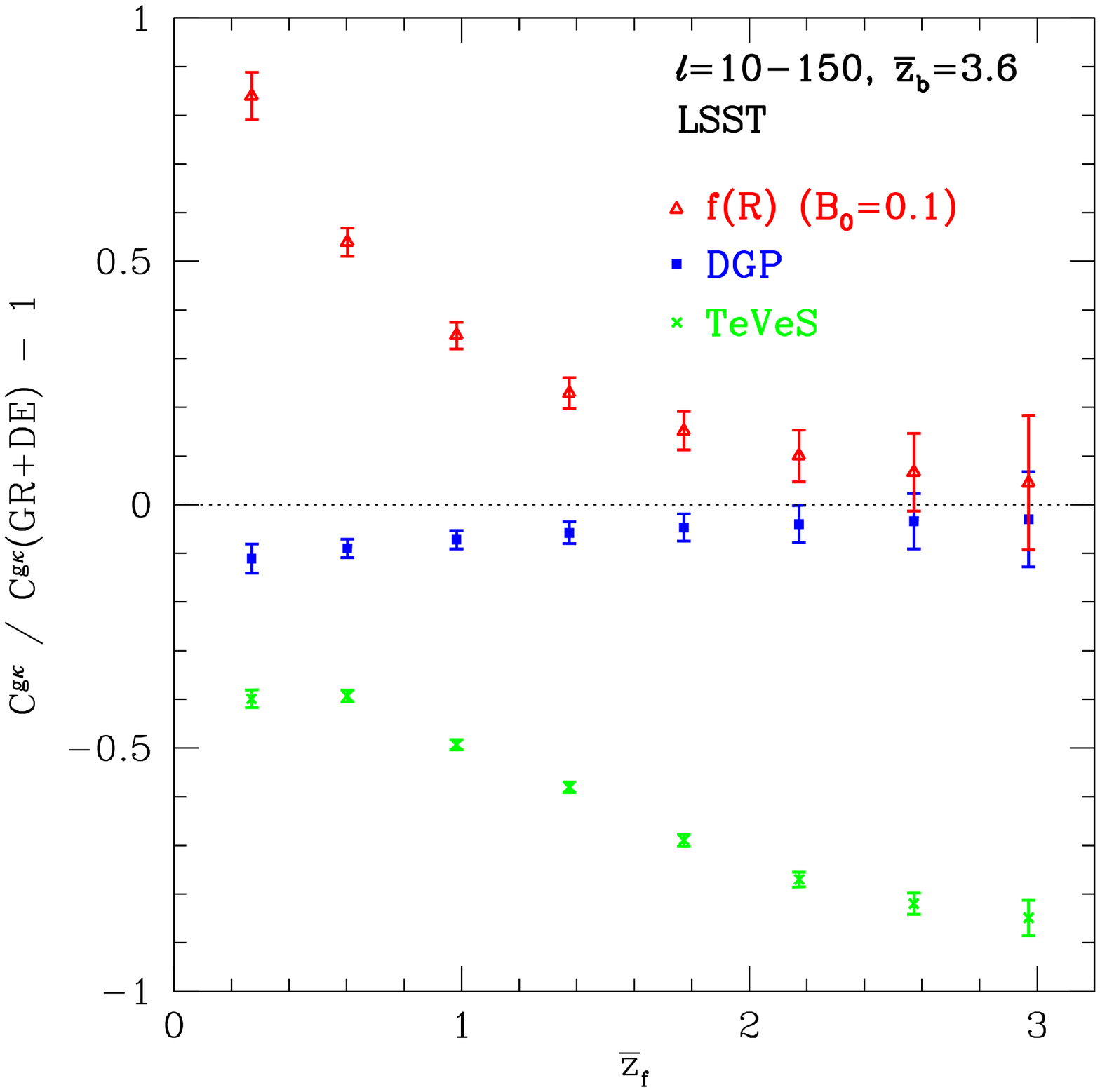}
\includegraphics[width=.48\textwidth]{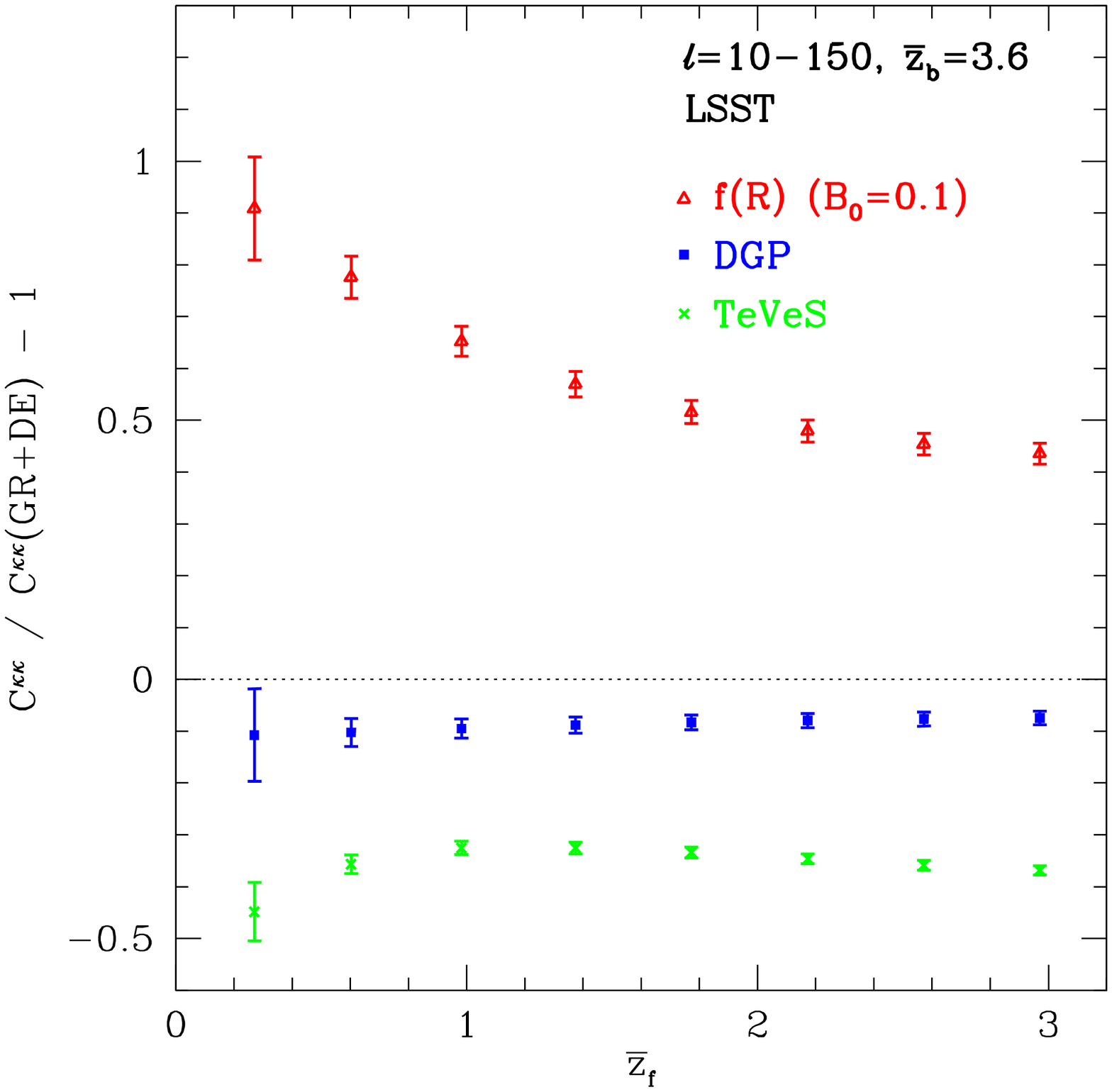}
\end{center}
\caption{
\textit{Left panel:} The relative difference of the galaxy-shear cross power
in modified gravity models, as
shown in \reffig{Cgk-vs-z} (right panel), but with statistical error
bars expected from LSST. Each point corresponds to an independent 
foreground galaxy redshift bin with $\Delta z = 0.4$; note that errors
are correlated as the same background galaxy sample is used for each point.
\textit{Right panel:} Same as the left panel, but for the shear
power spectrum for a fixed high-$z$ redshift
bin with $\zbar_b=3.6$ and the same foreground redshift bins as in the left panel.
\label{fig:C-vs-z-werr}}
\end{figure}

\subsection{Redshift evolution}

In order to study the redshift evolution of lensing correlations in
GR and modified gravity scenarios, we keep the background galaxy
bin ``B'' ($z=3-5$, median $\bar{z}_b=3.6$) fixed while considering
foreground redshift bins with various median redshifts 
as explained in \refsec{survey}.
We keep the width of the foreground redshift bins fixed at $\Delta z = 0.4$.
This is only a subset of the cross-correlations possible, which for our
purposes serves to show how the evolution of correlations differs in
modified gravity theories.

Let us begin with the galaxy-shear correlation, shown in \reffig{Cgk-vs-z}
(left panel) for a fixed $\l=100$ as a function of the median redshift
$\zbar_f$ of the foreground galaxies. The overall behavior as a function 
of $\zbar_f$ is due to the lensing weight function $W_L$, which peaks
at $z\sim 0.5$, corresponding to half of the distance to the lensed galaxies, 
of order $\chi(\zbar_b)$. The precise evolution with redshift however
depends on the evolution of the potentials $\propto (1+z)D_m(z)$, which
differs in modified gravity theories (right panel of \reffig{Cgk-vs-z}).
We again find the expected result that $\fR$ shows a larger lensing signal,
while DGP and TeVeS predict smaller lensing signals than the
corresponding GR+DE models. In all cases, we see
a significant redshift evolution of the deviations. 
In the case of $\fR$ and DGP, the deviations become smaller at higher $z$,
as GR becomes restored in the matter-dominated epoch in these models.
Thus, ratios
of correlations such as $C^{g\k}(\l;\zbar_f=0.5) / C^{g\k}(\l;\zbar_f=1.5)$
can serve as probes of gravity. A large part of the observational systematics
can be expected to drop out in such a ratio, as will the power spectrum
normalization $\sigma_8$. However, the different galaxy biases of the 
two foreground samples have to be taken into account. Alternatively,
one can measure ratios of $R^{g\k}$ [\refeq{Rgk}] instead of $C^{g\k}$,
where similar deviations from GR are seen.
It is apparent that at $\l=100$, non-linearities only become important 
for very low $z$ foreground galaxies in the galaxy-shear correlation.

The shear-shear correlation shows a complementary redshift evolution,
growing with the foreground galaxy redshfit $\zbar_f$, as the amount of 
mass along the line of sight as well as the geometrical lever arm increase 
with $\zbar_f$.
The deviations in the predicted redshift evolution of the shear-shear
correlation for modified gravity models are similar in sign and magnitude 
as those for the galaxy-shear correlation. Again, with its different redshift
evolution and set of observational systematics, the shear-shear correlation
can serve as a cross-check of the galaxy-shear correlation.

In \reffig{C-vs-z-werr} we show how precisely the redshift of evolution
of $C^{g\k}$ and $C^{\k\k}$ is expected to be measurable with a survey
like LSST. We assume a bin in $\l$ from 50--150, and consider foreground
redshift bins spaced by $\Delta z = 0.4$, so that these bins are
independent. Note however that the background redshift bin is the same
in all cases. It is clear that future surveys should be able to
resolve the redshift evolution of weak lensing correlations to high
precision, and distinguish the models discussed here based on this
evolution. This also holds when choosing a lower-$z$ background
redshift bin, e.g. $z=2-3$, although the redshift range usable for probing the
evolution of deviations from GR will be restricted to $z\lesssim 1.5$ 
in this case.

\begin{figure}[t]
\begin{center}
\includegraphics[width=.48\textwidth]{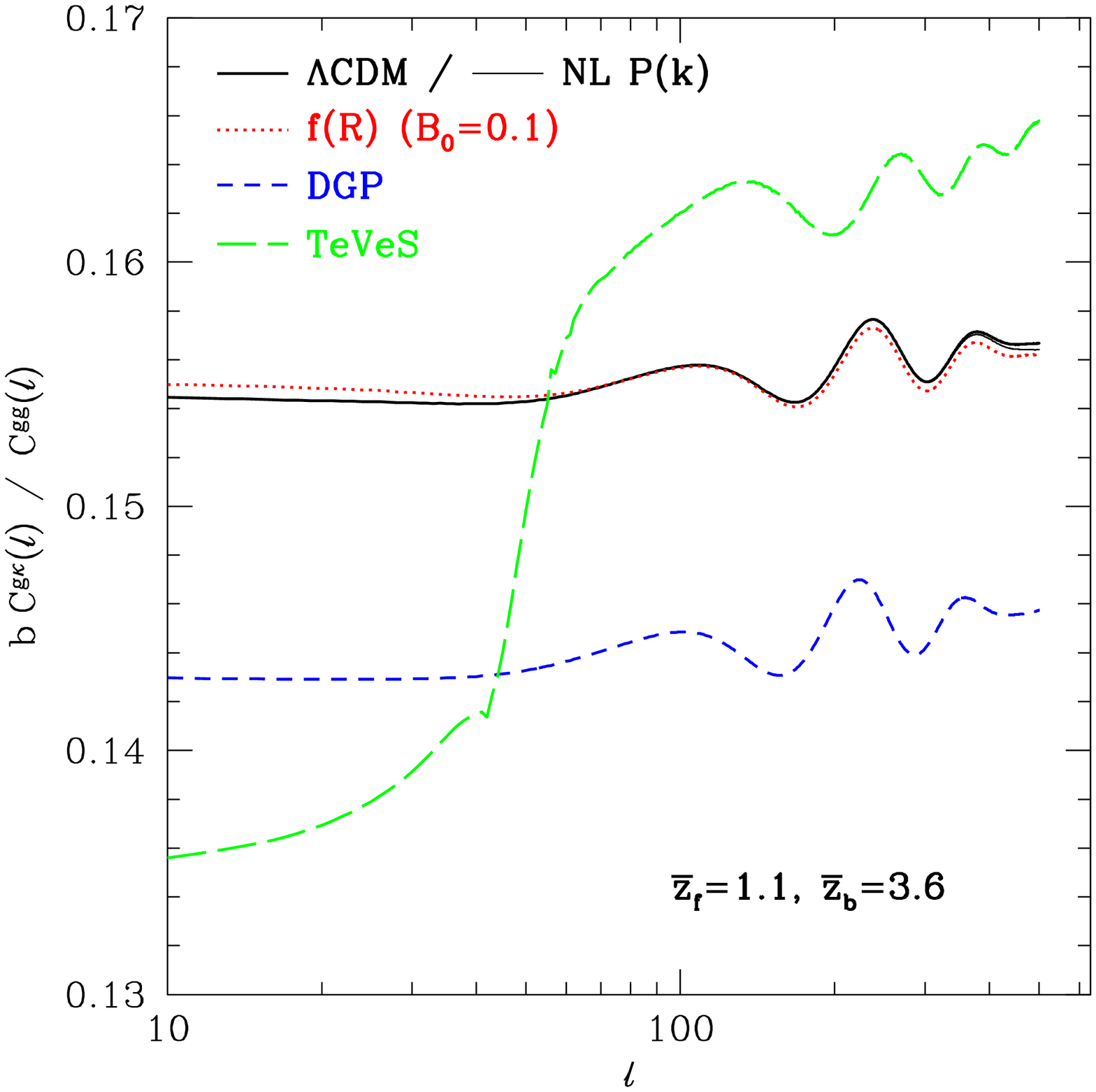}
\includegraphics[width=.48\textwidth]{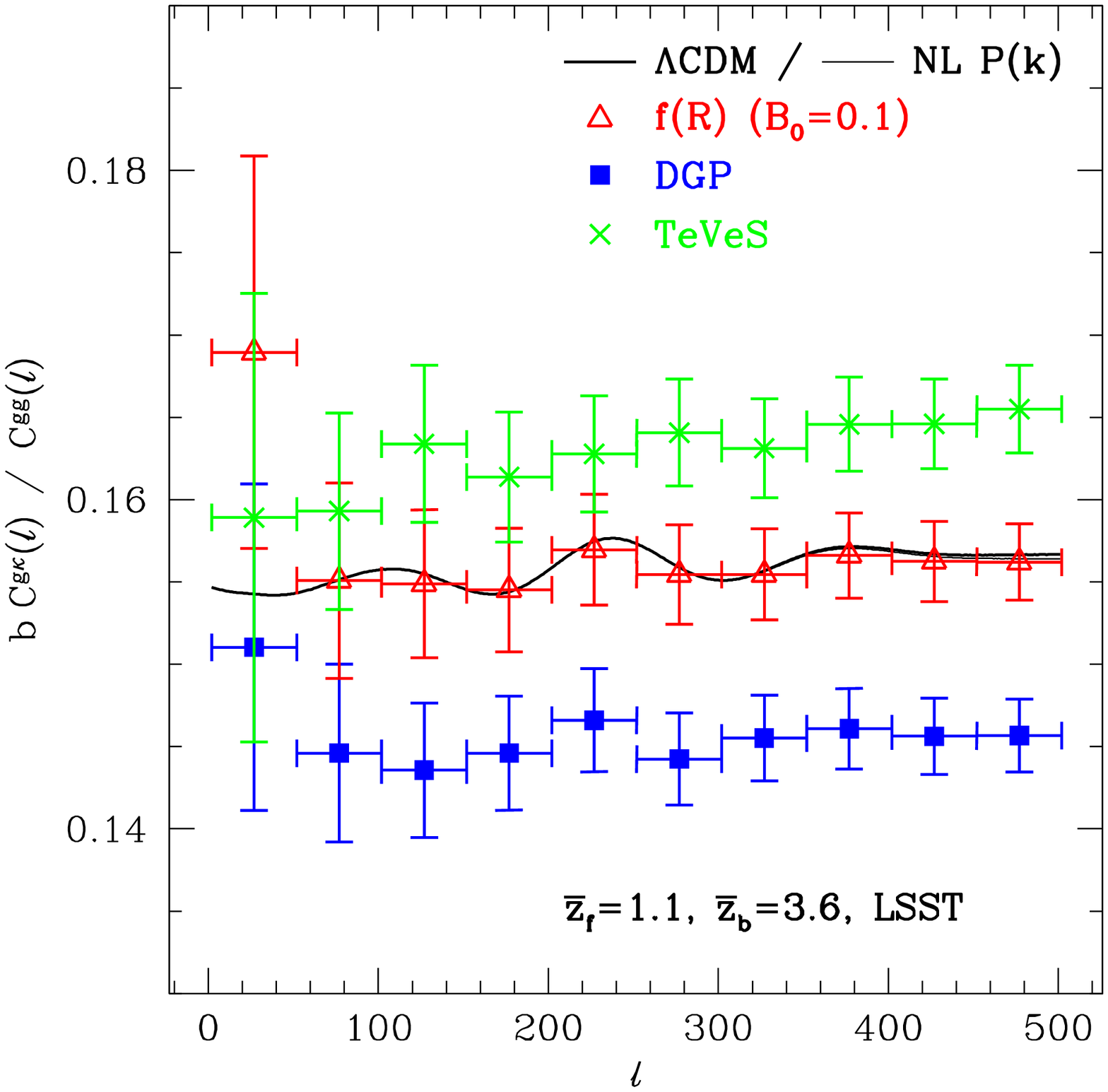}
\end{center}
\caption{\textit{Left panel:} The ``Poisson ratio'' of correlations
scaled by the bias, $b\:\P(\l)$ [\refeq{Poissonratio}],
for foreground galaxies with mean redshift
$\bar{z}_f=1.1$ and (sheared) background galaxies with $\bar{z}_b=3.6$, in 
$\Lambda$CDM and modified gravity theories.
\textit{Right panel:} The same as the left panel, in bins of $\Delta\l = 50$
and with statistical error bars expected from LSST.
\label{fig:Cgk-over-Cgg}}
\end{figure}

\subsection{Poisson ratio}
\label{sec:Pratio}

We now discuss forecasts for the measurement of the ``Poisson ratio''
$\P(\l) \equiv C^{g\k}(\l)/C^{gg}(\l)$ presented in \refsec{Poisson}. 
\reffig{Cgk-over-Cgg} shows $\P(\l)$ as a function of $\l$ for the foreground 
galaxy bin ``F'' ($\zbar_f=1.1$) and the background bin ``B'' ($\zbar_b=3.6$).
It is not completely scale-independent even in the case of $\Lambda$CDM due to
the finite width of the foreground galaxy bin ($\Delta z = 0.4$); 
in \refsec{Poisson} we had assumed a $\d$-distribution in redshift for 
the foreground galaxies. As expected from \reffig{DPhim}, TeVeS
shows the largest differences in $\P(\l)$. The modifications to the 
Poisson equation are very small for the assumed $\fR$ model, while the 
DGP model shows an offset in $\P(\l)$ due to the different $\Om$ adopted for
this model (\refsec{DGP}). Some of the generalizations of the DGP model
considered in \cite{Koyama2006} have a modified Poisson equation which
should also leave an observable signature in $\P(\l)$.

The right panel of \reffig{Cgk-over-Cgg} shows the statistical
error expected for LSST on the measurement of $\P(\l)$, in bins of
$\Delta\l = 50$. For simplicity, the errors on $C^{gg}$ and $C^{g\k}$ where
assumed to be independent. In this robust test of the Poisson equation, the 
present TeVeS model should leave an observational signature, as should
any modified gravity model that shows modifications to the 
Poisson equation on the order of 5\% or greater. Signatures of appreciable
Dark Energy density perturbations on scales less than $\sim$ Gpc should be 
detectable in $\P(\l)$ as well.

\begin{figure}[t]
\begin{center}
\includegraphics[width=.48\textwidth]{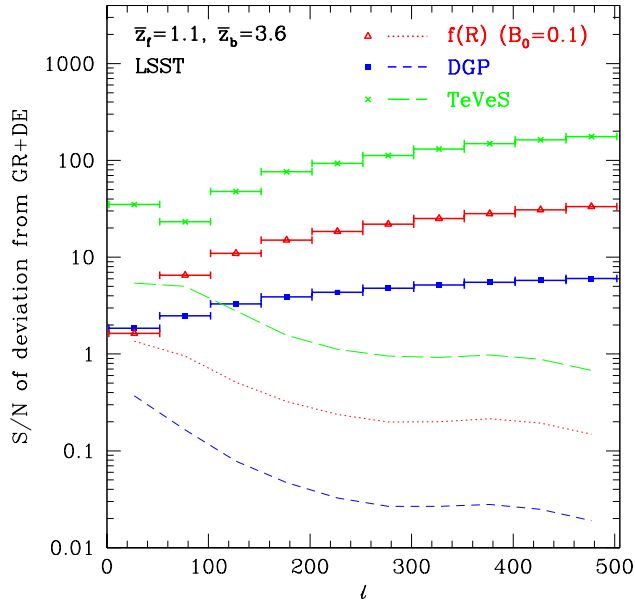}
\end{center}
\caption{Signal-to-noise $(C^{MG}(\l)-C^{GR+DE}(\l))/\Delta C^{MG}(\l)$ 
in $\l$ bins (points)
of the deviation of modified gravity (MG) predictions from GR+DE models with the
same expansion history, for the galaxy-shear correlation shown in \reffig{C-vs-l-werr}
(left panel). The lines show the corresponding signal to noise of the
deviation in the galaxy-CMB cross-correlation via the ISW effect, for the
same foreground galaxy redshift bin centered at $\bar{z}_f=1.1$.
\label{fig:SN-WL}}
\end{figure}

\section{Conclusions}
\label{sec:concl}

Uncovering the physics behind the accelerated expansion of the Universe is one of the 
most compelling open problems in astrophysics today. One fundamental question
to answer is whether the cause lies in an additional component in the 
energy budget of the Universe, or in a modification of General Relativity
on cosmological scales. By considering galaxy-shear and shear-shear 
correlations on large scales, we showed that weak lensing, with its 
ability to probe the scale dependence and redshift evolution of the 
cosmological gravitational potentials and their relation to matter,
can serve as a very sensitive probe in discerning between modified
gravity and Dark Energy. 
We focused on the effects of modified gravity on the growth
of structure, effectively assuming that the expansion history is very well
constrained through Supernovae Ia, the CMB, and baryon acoustic oscillation
measurements. In practice, distance and growth measures should be used
jointly to place constraints on modified gravity models.

Weak lensing is also a sensitive probe of the background
expansion history, both through geometry and the growth of structure. 
However, when using the growth of structure to infer the background expansion 
history (parametrized, e.g., by the Dark Energy equation of state $w$) ,
one relies on the validity of
General Relativity on cosmological scales: as we have shown here, modified
gravity affects the growth of structure and weak lensing observables
independently of the expansion history. Fortunately, it is possible to
isolate the dependence on the spacetime geometry of lensing observables 
via shear ratios \cite{JainTaylor,ZhangHuiStebbins}. These techniques will
not be affected by modifications to gravity. 

By cross-correlating foreground galaxies with the shear of background galaxies,
it is possible to probe the relation between
matter and potentials, i.e. the Poisson equation. For this purpose, we 
introduced the weak lensing correlation ratio $\P(\l)$ [\refeq{Poissonratio}], 
which is, in the limit of narrow
redshift bins, independent of growth effects and the matter power spectrum,
and thus isolates modifications to the matter-potential relation.
However, galaxy-shear correlations necessarily depend on 
the a priori unknown galaxy bias. On linear scales, the degeneracy with
the bias can presumably be broken by considering the non-trivial scale 
dependence of the modified gravity signatures; alternatively,
one can consider the reduced correlation, $R^{g\k}(\l)$ [\refeq{Rgk}].
In contrast, the shear-shear correlation is independent of any galaxy bias,
while it is more affected by the non-linear gravitational evolution at late 
times. We stress that any modification of gravity should leave a signature in 
\textit{both} galaxy-shear and shear-shear correlations, and the two methods 
with their different experimental systematics can serve as cross-checks of 
the results.

In this paper, we considered three viable modified gravity models which are,
in broad terms, consistent with current measurements of the CMB and the 
expansion history. We compared each model to a GR+DE scenario with 
identical expansion history, showing that weak lensing can break this
degeneracy even in the case of an extremely tightly constrained 
expansion history. Furthermore, all of these models make definite predictions
for weak lensing in the linear regime: the $f(R)$ model generically predicts a larger 
lensing signal than expected in GR; the DGP braneworld model predicts
a smaller lensing signal than a GR+DE model with the same expansion history;
and TeVeS predicts a weaker lensing signal with considerably modified 
scale dependence. In all cases, these are robust features of the underlying
model, linked to the non-zero difference between the cosmological potentials,
and modifications to the Poisson equation.

We showed that future wide-field weak lensing surveys, such as LSST and
the wide survey of SNAP, can detect deviations in weak lensing correlations
predicted by these three models with high significance. As an example,
\reffig{SN-WL} shows, as a function of $\l$, the signal-to-noise of the 
deviation of the modified gravity model predictions expected
for LSST, in the case of the galaxy-shear correlation.
For comparison, we also show the corresponding signal-to-noise expected 
for the galaxy-CMB cross-correlation induced by the late-time ISW effect,
which is clearly much smaller: weak 
lensing is considerably more powerful in distinguishing modified gravity 
models from Dark Energy models with the same expansion history.
As these conclusions hold for the three independent modified gravity
scenarios considered here, one might expect them to be valid in general
for gravity theories that differ significantly from GR on cosmological scales.
While constraints similar to those of LSST are expected from other proposed
surveys, such as the SNAP wide survey, these conclusions also hold for less 
demanding survey specifications.

Apart from relying on data from future experiments, progress in the 
theoretical understanding of modified gravity will be crucial in order
to improve on the sensitivity of probes of gravity. 
Using weak lensing to probe gravity on smaller scales
requires that the non-linear process of structure formation be understood
in modified gravity theories. Once this is the case, the amount of 
useful information for probing gravity grows dramatically, say by raising
$\l_{\rm max}$ from a few hundred to greater than 1000. 
In addition, one might hope that the degeneracy 
between modified gravity and general, non-smooth Dark Energy models present 
in the linear theory will be broken in the non-linear regime, at
least for a class of Dark Energy models restricted by physical constraints
on, e.g., the coupling to matter.

Finally, we point out that the information in weak lensing correlations
should be sufficient to place constraints on ``post-GR'' parameters
independently of an underlying modified gravity theory,
especially for the metric ratio $g(k,z)$ and the rescaled
gravitational constant $G_{\rm eff}$.
A detailed investigation of this, using all information contained
in galaxy and shear correlations, is left for future study.

\section*{Acknowledgments}

I am indebted to Scott Dodelson and Wayne Hu for many valuable suggestions 
and helpful input. I also thank Michele Liguori for providing the data
on the TeVeS model. This work was supported by the Kavli Institute 
for Cosmological Physics at the University of Chicago through grants 
NSF PHY-0114422 and NSF PHY-0551142 and an endowment from the Kavli 
Foundation and its founder Fred Kavli.

\bibliography{WLmodgrav}

\end{document}